\documentstyle[11pt,aaspp4]{article}

\load{\scriptsize}{\sc}

\input{psfig.tex}

\slugcomment{Submitted to Monthly Notices of the Royal Astronomical
 Society.}

\newcommand{\bfw}{\,\omega\hspace*{-2.65truemm}
 \omega\hspace*{-2.65truemm}\omega\hspace*{-2.65truemm}
 \omega\,}

\begin{document}


\title{RADIATIVE PRECESSION OF AN ISOLATED NEUTRON STAR}

\author{A. Melatos\altaffilmark{1,2,3}}
\affil{Theoretical Astrophysics, Mail Code 130--33, 
 California Institute of Technology, \\
 Pasadena, CA 91125 USA}

\altaffiltext{1}{Miller Fellow}
\altaffiltext{2}{New address: Department of Astronomy,
 601 Campbell Hall, University of California,
 Berkeley CA 94720 USA}
\altaffiltext{3}{E-mail: melatos@astraea.berkeley.edu}

\begin{abstract}
Euler's equations of motion are derived exactly for a 
rigid, triaxial, internally frictionless
neutron star spinning down electromagnetically 
{\em in vacuo}.
It is shown that the star precesses, but not freely:
its regular precession relative to the principal axes of inertia
couples to the
component of the radiation torque associated with 
the near-zone radiation fields
and is modified into an anharmonic wobble.
The wobble period $\tau_1$ typically satisfies
$\tau_1\lesssim 10^{-2}\tau_0$, where $\tau_0$ is
the braking time-scale;
the wobble amplitude 
evolves towards a constant non-zero value,
oscillates, or decreases to zero,
depending on the degree of oblateness or prolateness of
the star and its initial spin state;
and the (negative)
angular frequency derivative $\dot{\omega}$
oscillates as well, exhibiting quasi-periodic
spikes for triaxial stars of a particular figure.
In light of these properties,
a young, Crab-like pulsar ought to
display fractional changes of order unity
in the space of a few years in its
pulse profile, magnetic inclination angle,
and $\dot{\omega}$.
Such changes are not observed,
implying that the wobble is damped rapidly by internal
friction, if its amplitude
is initially large upon crystallization
of the stellar crust.
If the friction is localized in the inner and outer crusts,
the thermal luminosity of the neutron star
increases by a minimum amount
$\Delta L\approx 
 3\times10^{31}(\epsilon/10^{-12})
 (\omega / 10^3\,{\rm rad\,s^{-1}})^2 
 (\tau_{\rm d} / 1\,{\rm yr})^{-1}
 \,{\rm erg\,s^{-1}}$,
where $\epsilon$ is the ellipticity and $\tau_{\rm d}$
is the damping time-scale,
with the actual value of $\Delta L$ determined in part by
the thermal conduction time $\tau_{\rm cond}$.
The increased luminosity is potentially detectable as thermal
X-rays lasting for a time
$\approx {\rm max}(\tau_{\rm d},\tau_{\rm cond})$ 
following crystallization of the crust.
\end{abstract}

\clearpage


\section{INTRODUCTION \label{sec:tum1}}
Does the angular momentum vector of an isolated neutron star
change orientation as the star spins down?
Early attempts to answer this question focused on the
evolution of the angle $\alpha$ between the
star's rotation and magnetic axes,
which can be measured from pulsar polarization swings.
\markcite{dav70}Davis \& Goldstein (1970)
showed that $\alpha$ tends towards
zero on the braking time-scale
if the star is a rigid sphere or a fluid body in
hydrostatic equilibrium --- an unrealistic scenario
which leaves all except the youngest pulsars as
aligned rotators, contrary to observation.
\markcite{gol70}Goldreich (1970) observed that the
crystalline crust of a neutron star supports shear stresses,
thereby preventing a fraction of the hydrostatic bulge 
from aligning with the instantaneous rotation axis
and establishing
non-hydrostatic differences between the 
principal moments of inertia.
In the absence of internal friction,
such a triaxial star precesses
about its principal axis with a period 
that is short compared to the braking time-scale,
and the (fixed) angle between the
principal axis and magnetic axis determines whether the
precession amplitude (related to $\alpha$)
increases or decreases under the action of the
braking torque
(\markcite{gol70}Goldreich 1970).

In reality, a neutron star is not internally frictionless.
Elastic strain energy is dissipated in the crust
as the non-hydrostatic deformation discussed above
migrates around the star while it precesses
(\markcite{gol70}Goldreich 1970;
\markcite{cha71}Chau \& Henriksen 1971;
\markcite{mac74}Macy 1974).
There is also dissipation due to imperfect coupling
between the differentially rotating crust and
superfluid core
(\markcite{sha77}Shaham 1977;
\markcite{alp87}Alpar \& \"{O}gelman 1987;
\markcite{lin93}Link, Epstein \& Baym 1993;
\markcite{sed98}Sedrakian, Wasserman \& Cordes 1998).
Both types of friction damp any precession that is
initially present on a time-scale that is, theoretically,
much shorter than the precession period ---
a plausible explanation for
why the pulse profiles and polarization
characteristics of young pulsars
do not change secularly over several years
as one expects if the precession is undamped.
The only isolated neutron star unambiguously known to 
precess is PSR B1913$+$16
(\markcite{wei89}Weisberg, Romani \& Taylor 1989),
where general relativistic effects are responsible.
Tentative reports also exist
of oscillatory variations in Crab and
Vela timing residuals
(e.g.\ \markcite{lyn88b}Lyne, Pritchard \& Smith 1988;
\markcite{mcc90}McCulloch et al.\ 1990;
\markcite{cad97}\v{C}ade\v{z}, Gali\v{c}i\v{c} \& Calvani 1997),
and some authors have inferred changes in $\alpha$
on the braking time-scale
from braking-index measurements
(\markcite{all97}Allen \& Horvath 1997;
\markcite{lin97}Link \& Epstein 1997)
and the evolution of pulsar radio beam statistics
(\markcite{tau98}Tauris \& Manchester 1998).

The radiation torque acting on the neutron star has
been treated in an incomplete fashion in all of the above work.
The torque has two components: the familiar 
braking torque,
responsible for the secular spin-down of the star,
and a component associated with the
inertia of the near-zone radiation fields, 
sometimes misleadingly termed the `anomalous torque'
(\markcite{goo85}Good \& Ng 1985),
whose effect is to make a
spherical star precess about its magnetic axis.
In all analyses to date,
either the total radiation torque has been set to zero, 
in order to study free precession
(\markcite{mac74}Macy 1974;
\markcite{sha77}Shaham 1977;
\markcite{alp87}Alpar \& \"{O}gelman 1987),
or else the near-field component has been neglected, in the
belief that it exerts no significant influence
on the rotation
(\markcite{dav70}Davis \& Goldstein 1970;
\markcite{gol70}Goldreich 1970;
\markcite{cha71}Chau \& Henriksen 1971;
\markcite{sed98}Sedrakian et al.\ 1998).
A careful study of both torque components in the context
of a spherical star was carried out by
\markcite{goo85}Good \& Ng (1985) for a magnetic dipole,
a magnetic quadrupole, and
a hypothetical distribution of magnetospheric currents.
\markcite{cas98}Casini \& Montemayor (1998) recently
explored some effects of the near-field torque
on a composite body with a spherically symmetric crust 
coupled to a spherical core
(cf.\ \markcite{dec80}de Campli 1980).

In this paper, we demonstrate that the near-field
component of the radiation torque strongly influences
the rotation of an internally frictionless neutron star.
In Section \ref{sec:tum2}, we derive and solve Euler's 
equations of motion for a rigid, triaxial magnet
and show that the regular precession relative to
the principal axes of inertia couples to the near-field
torque, causing the star to wobble
anharmonically. Certain potentially observable properties
of the wobble are explored in Section \ref{sec:tum3},
including its period and amplitude and the slow
evolution of $\alpha$ and the angular frequency derivative
$\dot{\omega}$.
The results are applied to pulsar timing and polarization
observations, and to the internal structure of young neutron
stars, in Section \ref{sec:tum4}.

\section{ROTATION OF A RIGID, TRIAXIAL NEUTRON STAR 
 \label{sec:tum2}}
In this section, the rotation of a rigid, triaxial body
with an embedded magnetic dipole is treated analytically.
Euler's equations of motion are written down
in Section \ref{sec:tum2a}.
Two key elements of the motion --- 
the torque-driven radiative precession, and the inertial
free precession --- are discussed in Sections \ref{sec:tum2b1}
and \ref{sec:tum2b2} respectively and their time-scales
are identified.
Euler's equations are then solved approximately in
Section \ref{sec:tum2c} by a time-averaging technique
for the special case of a biaxial star.
The similarities and differences between this treatment
and previous work are noted in Section \ref{sec:tum2d}.

\subsection{Euler's Equations \label{sec:tum2a}}
Consider a rigid, triaxial star with principal axes
${\bf e}_1$, ${\bf e}_2$ and ${\bf e}_3$,
corresponding principal moments of inertia
$I_1$, $I_2$ and $I_3$,
ellipticities $\epsilon=(I_3-I_1)/I_1$
and $\epsilon'=(I_2-I_1)/I_1$,
and average radius $r_0$. 
We assume that the stellar magnetic field is dipolar
and fixed in the star, and we restrict 
the magnetic axis ${\bf m}$
to lie in the plane spanned by 
${\bf e}_1$ and ${\bf e}_3$,
at an angle $\chi$ to ${\bf e}_3$;
this entails a slight loss of generality.
We also assume that the star is internally frictionless 
and that it rotates {\em in vacuo},
so that it is unaffected by magnetospheric currents;
refer to
\markcite{gol70}Goldreich (1970) and 
\markcite{mel97}Melatos (1997) 
for justifications of the latter assumption.
It is important to keep in mind that,
although we idealize the star as a rigid body for simplicity,
in reality the ellipsoid of inertia is determined by an
equilibrium between elastic and hydromagnetic forces,
as discussed in Section \ref{sec:tum2b2}.

In Appendix \ref{sec:tumappa}, we evaluate the
radiation torque acting on the star from
the electromagnetic fields generated by a magnetized,
conducting sphere rotating {\em in vacuo}
(\markcite{deu55}Deutsch 1955;
\markcite{mel97}Melatos 1997);
it is assumed that the star's triaxiality
can be neglected when calculating the fields
and torque.
Upon resolving the torque into components along
the principal axes, we arrive at Euler's equations
of motion,
\begin{eqnarray}
 \dot{u}_1
 & = &
 (\epsilon' -\epsilon) u_2 u_3 +
 (\omega_0 \tau_0)^{-1} \cos\chi
 \left[
  u^2 F(x_0) (-u_1 \cos\chi + u_3 \sin\chi)
 \right. \nonumber \\
 & & 
  \left.
  \phantom{ u^2 }
  + u G(x_0) u_2 (u_1 \sin\chi + u_3 \cos\chi) 
 \right],
 \label{eq:tum2}
 \\
 (1 + \epsilon') \dot{u}_2
 & = &
 \epsilon u_1 u_3 +
 (\omega_0 \tau_0)^{-1}
 \left[
  -u^2 F(x_0) u_2
 \right. \nonumber \\
 & & 
  \left.
  \phantom{ u^2 }
  + u G(x_0) (-u_1 \cos\chi + u_3 \sin\chi)
   (u_1 \sin\chi + u_3 \cos\chi) 
 \right],
 \label{eq:tum3}
 \\
 (1 + \epsilon) \dot{u}_3
 & = &
 - \epsilon' u_1 u_2
 - (\omega_0 \tau_0)^{-1} \sin\chi
 \left[
  u^2 F(x_0) (-u_1 \cos\chi + u_3 \sin\chi)
 \right. \nonumber \\
 & & 
  \left.
  \phantom{ u^2 }
  + u G(x_0) u_2 (u_1 \sin\chi + u_3 \cos\chi) 
 \right],
 \label{eq:tum4}
\end{eqnarray}
with
\begin{equation}
 F(x_0)
 =
 \frac{x_0^4}{5(x_0^6 - 3x_0^4 + 36)}
 +
 \frac{1}{3(x_0^2 + 1)},
 \label{eq:tum5}
\end{equation}
\begin{equation}
 G(x_0)
 =
 \frac{3(x_0^2 + 6)}{5x_0 (x_0^6 - 3x_0^4 + 36)}
 +
 \frac{3 - 2x_0^2}{15 x_0 (x_0^2 + 1)}.
 \label{eq:tum6}
\end{equation}
In equations (\ref{eq:tum2}) to (\ref{eq:tum6}),
$\omega_0$ is the magnitude of the angular velocity vector
$\bfw (t)$ at time $t=0$,
we define dimensionless variables
${\bf u}=\bfw/\omega_0$
and
$x_0 = (r_0 \omega_0 / c) u$,
an overdot denotes differentiation with respect to 
the dimensionless time coordinate $s=\omega_0 t$,
and
\begin{equation}
 \tau_0 =
 \frac{\mu_0 c^3 I_1}
  {2\pi B_0^2 r_0^6 \omega_0^2}
 \label{eq:tum1}
\end{equation}
is the characteristic braking time-scale at $t=0$,
in terms of the magnetic field strength $B_0$
at the magnetic poles
(\markcite{deu55}Deutsch 1955).

The form factors $F(x_0)$ and $G(x_0)$ reflect the
structure of the near-zone radiation fields through their
dependences on $x_0$. 
The familiar braking torque, which is responsible
for the secular spin-down of the star, is associated
with terms proportional to $F(x_0)$, 
whereas the near-field torque
discussed in Section \ref{sec:tum1} is associated
with terms proportional to $G(x_0)$.
In most applications, $r_0$ is taken to
be the stellar radius $R$ (cf.\ Section \ref{sec:tum3f}
and \markcite{kab81}Kaburaki 1981, \markcite{mel97}Melatos 1997), 
yielding $x_0\ll 1$, 
$F(x_0)=1/3$ and $G(x_0)=3/10x_0$. 
The near-field torque
is therefore much greater than the braking torque
in this regime and acts on a commensurately shorter
time-scale.
Equations (\ref{eq:tum5}) and (\ref{eq:tum6})
differ from the expressions 
$F(x_0)=1/3$ and $G(x_0)=1/2x_0$
found in previous works
(\markcite{dav70}Davis \& Goldstein 1970;
\markcite{gol70}Goldreich 1970),
partly because the treatment in this paper
is not restricted to $x_0 \ll 1$,
and partly because we
model the star's internal magnetization
in a slightly different way,
as explained in Appendix \ref{sec:tumappa}.

\subsection{Radiative Precession
 \label{sec:tum2b1}}
The near-field component of the radiation torque causes the star
to precess and nutate about its magnetic axis. We call this
motion `radiative precession'.
To understand its origin, consider the simple special case
of a spherical star ($\epsilon=\epsilon'=0$).
In the regime $x_0\ll 1$, where the form factor
$uG(x_0)\approx 3c/10 r_0 \omega_0$
is independent of $s$,
Euler's equations (\ref{eq:tum2}) to (\ref{eq:tum4})
have the exact solution
(cf.\ \markcite{dav70}Davis \& Goldstein 1970)
\begin{eqnarray}
 u_1
 & = &
 u_{1,0} u_{3,0}
 \left\{
  \exp\left[
   \frac{2 u_{3,0}^2 F(x_0) s}{\omega_0\tau_0} 
  \right]
  - u_{1,0}^2 
 \right\}^{-1/2}
 \cos\left[
   \frac{u_{3,0} uG(x_0) s }{\omega_0 \tau_0}
  \right],
 \label{eq:tum12}
 \\
 u_2
 & = &
 u_{1,0} u_{3,0}
 \left\{
  \exp\left[ 
   \frac{2 u_{3,0}^2 F(x_0) s}{\omega_0\tau_0} 
  \right]
  - u_{1,0}^2 
 \right\}^{-1/2}
 \sin\left[
   \frac{u_{3,0} uG(x_0) s }{\omega_0 \tau_0}
  \right],
 \label{eq:tum13}
 \\
 u_3
 & = &
 u_{3,0}~,
 \label{eq:tum14}
\end{eqnarray}
for the initial conditions 
$u_1=u_{1,0}$, $u_2=0$, $u_3=u_{3,0}\neq 0$ 
at $s=0$.\footnote{See \markcite{dav70}Davis \& Goldstein (1970) 
for a discussion of the singular special case $u_{3,0}=0$.}
A spherical
star therefore precesses harmonically about ${\bf m}$
with period
\begin{equation}
 \tau_1
 =
 \left(
 \frac{20\pi r_0 \omega_0}{3 u_{3,0} c}
 \right)
 \tau_0
 \ll
 \tau_0~,
 \label{eq:tum15}
\end{equation}
and the precession amplitude decays exponentially
on the time-scale $3\tau_0/u_{3,0}^2$,
as $\bfw$ aligns with ${\bf m}$.
Taking $r_0=R=10\,{\rm km}$
(cf.\ Section \ref{sec:tum3f}) and 
$I_1=1\times10^{38}\,{\rm kg\,m^2}$, 
we obtain numerically
\begin{equation}
 \tau_1
  =
 4 \times 10^{13}
 \left( \frac{B_0}{10^{8}\,{\rm T}} \right)^{-2}
 \left( \frac{\omega}{1\,{\rm rad\,s^{-1}}} \right)^{-1}
 {\rm s}~,
 \label{eq:tum17a}
\end{equation}
which is to be compared with
\begin{equation}
 \tau_0
  =
 5 \times 10^{16}
 \left( \frac{B_0}{10^{8}\,{\rm T}} \right)^{-2}
 \left( \frac{\omega}{1\,{\rm rad\,s^{-1}}} \right)^{-2}
 {\rm s}~.
 \label{eq:tum17b}
\end{equation}

Radiative precession occurs due to the asymmetric inertia
of the near-zone radiation fields
of a rotating magnetic dipole
(\markcite{gol70}Goldreich 1970):
the electromagnetic energy density ${\cal E}$ 
is greater at the magnetic poles
(${\cal E}\approx B_0^2/2\mu_0$) 
than at the magnetic equator
(${\cal E}\approx B_0^2/8\mu_0$),
translating into a fractional distortion
$\epsilon^{\rm rad}\approx 
 (\delta{\cal E} / c^2) r_0^5 /  I_1$
of the moment of inertia about ${\bf m}$.
The effect depends only on the radiation fields
outside the star, which in turn are determined completely
by the magnetic field at the stellar surface and the
property that the star is a good conductor
(\markcite{deu55}Deutsch 1955).
The magnetic field inside the star does not influence
the radiative precession.

The near-field torque contributes terms proportional
to $u_2 u_3$ and $u_1 u_3$ in (\ref{eq:tum2}) and
(\ref{eq:tum3}) respectively and therefore adds to,
or subtracts from, similar terms arising from material
distortions (i.e.\ $\epsilon$ and $\epsilon'$).
Thus a biaxial star
($\epsilon' = 0$) with $\chi=0$ and
$\epsilon=\epsilon^{\rm rad}$ does not precess at all,
because the terms 
$(\epsilon'-\epsilon) u_2 u_3$ and
$\epsilon u_1 u_3$ are cancelled out exactly by the 
near-field torque.

\subsection{Free Precession
 \label{sec:tum2b2}}
In the absence of the radiation torque, a triaxial neutron
star precesses relative to its principal axes of inertia
on a time-scale $\tau_2 = 2\pi / \epsilon\omega$,
where $\epsilon$ is the non-hydrostatic ellipticity.
A variety of non-hydrostatic mechanisms, 
many with geological analogues 
(\markcite{lam80}Lambeck 1980), 
combine to deform the stellar mass distribution.
We concentrate on elastic and magnetic deformations
in this paper.

The crystalline stellar crust supports shear
stresses which prevent a fraction of the hydrostatic bulge 
from aligning with the instantaneous rotation axis.
For a crust with uniform shear modulus $\mu$,
one finds
$\epsilon^{\rm cr} =
 5\tilde{\mu}\omega^2 R^3 / 
 4(1+\tilde{\mu})GM$,
with $\tilde{\mu}=38\pi \mu R^4/ 3 G M^2$,
where $M$ is the mass of the star
(\markcite{gol70}Goldreich 1970;
 \markcite{lam80}Lambeck 1980, p.\ 42).
This elastic deformation yields a precession period
\begin{equation}
 \tau_2^{\rm cr}
  =
 1 \times 10^{14}
 \left( \frac{\mu}{10^{28}\,{\rm N\,m^{-2}}} \right)^{-1}
 \left( \frac{\omega}{1\,{\rm rad\,s^{-1}}} \right)^{-3}
 {\rm s}
 \label{eq:tum17c}
\end{equation}
for neutron star parameters.
If the crust is structured as a Coulomb lattice,
one has $\tilde{\mu}\lesssim 10^{-4}$
and the non-hydrostatic fraction of the bulge is small;
however, the exact value of $\tilde{\mu}$ is uncertain 
and may be much less than this upper bound.

The magnetic field inside the star creates an additional
deformation because non-radial field gradients (e.g.\ between
the poles and equator if the field is a dipole) support
non-radial matter-density gradients in hydromagnetic
equilibrium. (This is not related in any way
to the electromagnetic inertia of the external radiation
fields discussed in Section \ref{sec:tum2b1}.)
The geometry and magnitude of the deformation is difficult to
estimate, because little is known about the structure of
the internal magnetic field.
\markcite{tho93}Thompson \& Duncan (1993)
argued that, if the internal field is generated after collapse
in a convective dynamo, then it must be organized into randomly
oriented loops $\sim 1\,{\rm km}$ in size, each with field strength
$\sim 10^{11}\,{\rm T}$; the neutron star rotates too slowly
(Rossby number ${\rm Ro} \gtrsim 10$) to establish a coherent
toroidal field at the base of the convection zone
(cf.\ the Solar dynamo).
On the other hand, if the internal field is generated before
collapse in the progenitor star (e.g.\ in the convective
outer envelope or hydrogen-burning core), a large toroidal
field can grow (${\rm Ro} \lesssim 0.1$).
\markcite{bla83}Blandford, Applegate \& Hernquist (1983) examined toroidal
field generation by a thermoelectric dynamo.
Several authors have raised the possibility of a very strong
internal field ($\gtrsim 10^{10}\,{\rm T} \gg B_0$)
in diverse contexts, including off-centred-dipole theories
of the pulsar death line
(\markcite{aro98}Arons 1998),
thermally regulated resurrection of a buried field 
(\markcite{mus96}Muslimov \& Page 1996),
Ohmic decay in an anisotropically conducting core 
(\markcite{hae90}Haensel, Urpin \& Yakovlev 1990),
crust-core coupling in Vela glitches
(\markcite{abn96}Abney, Epstein \& Olinto 1996; 
 cf.\ \markcite{eas79}Easson 1979),
and the effect of a virial field ($10^{14}\,{\rm T}$)
on modified Urca cooling
(\markcite{yua98}Yuan \& Zhang 1998)
and the quark-hadron equation of state
(\markcite{pal98}Pal, Bandyopadhyay \& Chakrabarty 1998).

Assuming that the internal magnetic field
is at least as strong as the surface field and roughly dipolar,
we find that the hydromagnetic deformation satisfies
$\epsilon^{\rm mag} \gtrsim
 \epsilon^{\rm rad} c^2 / c_{\rm s}^2$,
where $c_{\rm s}$ is the isothermal sound speed
($=3^{-1/2}c$ in a relativistic star);
in other words, $\epsilon^{\rm mag}$ exceeds
$\epsilon^{\rm rad}$ at least by a factor of order the ratio
of the stellar to Schwarzschild radii.
This yields an upper bound on the precession period 
$\tau_2^{\rm mag}$ associated with the hydromagnetic 
deformation given by
\begin{equation}
 \tau_2^{\rm mag} \leq
 1\times 10^{13}
 \left( \frac{B_0}{10^{8}\,{\rm T}} \right)^{-2}
 \left( \frac{\omega}{1\,{\rm rad\,s^{-1}}} \right)^{-1}
 {\rm s}~.
 \label{eq:tum10}
\end{equation}
Note that the combined elasto-hydromagnetic deformation is
triaxial in general. Biaxiality ($\epsilon'=0$) is a good
approximation only when one has 
$\epsilon^{\rm cr} \ll \epsilon^{\rm mag}$
(or else $\epsilon^{\rm cr} \gg \epsilon^{\rm mag}$)
and the internal magnetic field (or crust) is symmetric about
a unique axis.

Equations (\ref{eq:tum17a}), (\ref{eq:tum17c}) and
(\ref{eq:tum10}) reveal that, as a rule,
{\em the periods of the radiative precession and free precession
are comparable}.
One has $\tau_1\sim\tau_2^{\rm mag}$,
if the internal and surface magnetic fields are of similar
magnitude, and sometimes
$\tau_1\sim\tau_2^{\rm cr}$ as well,
e.g.\ for a one-second pulsar with 
$\mu=10^{28}\,{\rm N\,m^{-2}}$ and 
$B_0=4\times 10^{8}\,{\rm T}$.
In general, therefore,
the `free' precession is not free at all;
rather,
it couples to the near-field component of the radiation torque,
and its character is modified significantly as a result.
This coupling has been overlooked in the literature 
on aspherical rotators to date
(\markcite{gol70}Goldreich 1970;
\markcite{cha71}Chau \& Henriksen 1971;
\markcite{mac74}Macy 1974;
\markcite{sha77}Shaham 1977;
\markcite{alp87}Alpar \& \"{O}gelman 1987;
\markcite{sed98}Sedrakian et al.\ 1998),
and the remainder of this paper is devoted to 
exploring its consequences.
It can only be neglected under certain circumstances,
e.g.\ when the internal magnetic field is strong and the
free precession is therefore fast 
($\tau_2^{\rm mag}\lesssim 0.05\tau_1$; 
 see Section \ref{sec:tum3b}).

\subsection{Separating the Braking and Precession Time-Scales:
 An Approximate Solution of Euler's Equations
 \label{sec:tum2c}}
Euler's equations (\ref{eq:tum2}) to (\ref{eq:tum4})
can be solved approximately by averaging over the
precession period, exploiting the fact that both $\tau_1$
and $\tau_2$ are small compared to
$\tau_0$ (except in the regime $x_0\sim 1$;
see Section \ref{sec:tum3f}).
The analysis parallels that by
\markcite{gol70}Goldreich (1970), with one crucial
difference: 
we account fully for
the radiative precession in what follows, whereas
\markcite{gol70}Goldreich (1970)
artificially suppressed it by averaging the braking
torque {\em and near-field torque}
over the free precession period.

Let us restrict attention to a biaxial star ($\epsilon'=0$)
for the sake of simplicity.
When the `slow' braking terms in Euler's equations,
proportional to $(\omega_0\tau_0)^{-1} F(x_0)$,
are neglected relative to `fast' terms, proportional to
$\epsilon$ and $(\omega_0\tau_0)^{-1} u G(x_0)$,
the equations reduce to the zeroth-order system
\begin{eqnarray}
 \dot{u}_1 
 & = &
 \epsilon\beta^{-1} u_2
 [ u_1\sin\chi\cos\chi -
   (\beta-\cos^2\chi) u_3 ],
 \label{eq:tum18}
 \\
 \dot{u}_2
 & = &
 \epsilon\beta^{-1}
 [ (u_3^2-u_1^2)\sin\chi\cos\chi +
   (\beta-\cos^2\chi+\sin^2\chi) u_1 u_3 ],
 \label{eq:tum19}
 \\
 \dot{u}_3 
 & = &
 - \epsilon\beta^{-1} u_2
 ( u_1\sin\chi + u_3\cos\chi )
 \sin\chi,
 \label{eq:tum20}
\end{eqnarray}
with
\begin{equation}
 \beta
 =
 \frac{\epsilon\omega_0\tau_0}{uG(x_0)}~.
 \label{eq:tum21}
\end{equation}
The parameter $\beta$ is independent of $s$ 
in the regime $x_0\ll 1$ and is positive or negative
according to whether the star is oblate or prolate.
Upon multiplying 
(\ref{eq:tum18}), (\ref{eq:tum19}) and (\ref{eq:tum20})
by $u_1$, $u_2$ and $u_3$ respectively, 
we find that
\begin{equation}
 \eta=u_1^2 + u_2^2 + u_3^2
 \label{eq:tum22}
\end{equation}
is a constant of the motion; in other words, the total
angular momentum of the star is unaffected by the
near-field component of the radiation torque.
Furthermore, dividing 
(\ref{eq:tum18}) by (\ref{eq:tum20})
yields a second constant of the motion
\begin{equation}
 \gamma
 =
 (u_1\sin\chi + u_3\cos\chi)^2 -
 \beta u_3^2~,
 \label{eq:tum23}
\end{equation}
which loosely measures the difference 
between the precession and magnetic inclination angles.
One can solve 
(\ref{eq:tum19}), (\ref{eq:tum22}) and (\ref{eq:tum23})
simultaneously to obtain a first-order
differential equation for $u_2$,
solvable by quadrature,
whose solution is exactly periodic but anharmonic
in general.
For the illustrative special case 
$\chi=90\arcdeg$, $\beta>0$ 
calculated in Appendix \ref{sec:tumappb},
we find
\begin{eqnarray}
 u_1 
 & = &
 \left(
  \frac{\gamma+\eta\beta}{\beta+1}
 \right)^{1/2}
 {\rm cn}
 \left[ \epsilon
  \left( \frac{\eta-\gamma}{\beta} \right)^{1/2} s
  + \Phi
 \right],
 \label{eq:tum24}
 \\
 u_2 
 & = &
 \left(
  \frac{\gamma+\eta\beta}{\beta}
 \right)^{1/2}
 {\rm sn}
 \left[ \epsilon
  \left( \frac{\eta-\gamma}{\beta} \right)^{1/2} s
  + \Phi
 \right],
 \label{eq:tum25}
 \\
 u_3 
 & = &
 \left(
  \frac{\eta-\gamma}{\beta+1}
 \right)^{1/2}
 {\rm dn}
 \left[ \epsilon
  \left( \frac{\eta-\gamma}{\beta} \right)^{1/2} s
  + \Phi
 \right],
 \label{eq:tum26}
\end{eqnarray}
where sn, cn and dn are Jacobian elliptic functions
with modulus 
$k^2=(\gamma+\eta\beta)/(\eta-\gamma)\beta$.
Note that the phase $\Phi$ of the oscillation is a third
constant of the motion, related to  $u_1$, $u_2$ and $u_3$
in a complicated way.

When terms proportional to 
$(\omega_0\tau_0)^{-1} F(x_0)$
are restored to Euler's equations,
the above solutions remain approximately valid,
but the constants of the motion are converted into
slowly varying functions of $s$, viz.,
$\eta=\eta(s)$, $\gamma=\gamma(s)$ and $\Phi=\Phi(s)$.
The slow variation occurs on the braking
time-scale $\tau_0$, which is long compared to
$\tau_1\sim\tau_2\sim\epsilon^{-1}$.
Analytic expressions for $\dot{\eta}$, $\dot{\gamma}$
and $\dot{\Phi}$ are obtained by substituting
(\ref{eq:tum24}), (\ref{eq:tum25}) and (\ref{eq:tum26})
into
(\ref{eq:tum2}), (\ref{eq:tum3}) and (\ref{eq:tum4})
and averaging over $\Phi$.
Results for $\chi=90\arcdeg$ are given
for reference in Appendix \ref{sec:tumappb}.

\subsection{Comparison with Previous Work
 \label{sec:tum2d}}
Equations (\ref{eq:tum2}) to (\ref{eq:tum4})
reduce to the equations solved by
\markcite{dav70}Davis \& Goldstein (1970)
for the special case of a spherical star
($\epsilon=\epsilon'=0$, and hence $\chi=0$ without loss
of generality),
and to those solved by
\markcite{gol70}Goldreich (1970) 
in the regime $\tau_2\ll\tau_1$ where the
free precession is much faster than the
radiative precession.
The latter regime
corresponds to a large elasto-hydromagnetic
deformation, such as when the internal magnetic field is 
much stronger than the surface field
(see Section \ref{sec:tum2b2}),
but it is not fully general.
In contrast, the solutions presented in Section \ref{sec:tum2c}
and the numerical results  in Section 3 address the general
problem where $\tau_1/\tau_2$ is arbitrary,
including the regime $\tau_1\approx \tau_2$ where the
free and radiative precessions couple together.
Previous authors have discussed
the physical origins of $\tau_1$ and $\tau_2$
(\markcite{gol70}Goldreich 1970; 
\markcite{cha71}Chau \& Henriksen 1971;
\markcite{dec80}de Campli 1980).

Our analysis cannot be compared directly
with previous work treating the star as anelastic
(e.g.\ \markcite{mac74}Macy 1974)
or with models featuring a core and crust coupled
together
(\markcite{sha77}Shaham 1977;
\markcite{dec80}de Campli 1980;
\markcite{alp87}Alpar \& \"{O}gelman 1987;
\markcite{cas98}Casini \& Montemayor 1998;
\markcite{sed98}Sedrakian et al.\ 1998).

\section{PROPERTIES OF THE ROTATION
 \label{sec:tum3}}
The motion analyzed in Section \ref{sec:tum2}
is characterized by several properties of potential
observational significance which we now investigate,
including the precession period
(Section \ref{sec:tum3a}) and amplitude
(Section \ref{sec:tum3b}),
the evolution of the magnetic inclination
(Section \ref{sec:tum3c}),
the different behaviour of oblate and prolate stars
(Section \ref{sec:tum3d}),
the phenomenon of `pseudo-glitches' in $\dot{\omega}$
(Section \ref{sec:tum3e}),
and the effect of a corotating magnetosphere
(Section \ref{sec:tum3f}).
In what follows, we assume that
the orientation of $\bfw$ is arbitrary
at the time when the neutron star first crystallizes into
an object with a rigid crust
spinning down electromagnetically.
In other words, $u_{1,0}$, $u_{2,0}$ and $u_{3,0}$
are assumed to be comparable; we do not choose a privileged
initial spin state where
$\bfw$ is parallel to one of the principal axes.
This assumption is important because
in Section \ref{sec:tum4} we present strong
observational evidence that isolated pulsars evolve 
rapidly towards
a stable state of this sort, probably under the
action of internal friction.
Consequently, the properties investigated below
characterize neutron stars early in their lives.

The results in this section are mainly presented in the
context of a biaxial star ($\epsilon'=0$) for two reasons.
First, we find numerically that the rotation properties
of biaxial and triaxial stars are qualitatively alike,
except for pseudo-glitches (Section \ref{sec:tum3e})
which are an intrinsically triaxial phenomenon.
Second, the aim of this section is to illustrate those
aspects of the dynamics that are observationally relevant;
an exhaustive quantitative survey of the rotation of a triaxial
magnet lies outside the scope of this paper.
Formally speaking, however, biaxiality is a good approximation 
only as long as one has
$\epsilon^{\rm cr} \ll \epsilon^{\rm mag}$
(or else $\epsilon^{\rm cr} \gg \epsilon^{\rm mag}$)
and the internal magnetic field (or crust)
is symmetric about a unique axis.

\subsection{Precession Period
 \label{sec:tum3a}}
The period $\Delta s$ of the anharmonic precession
is typically
$\sim|\beta|^{1/2}|\epsilon|^{-1}$,
in units of $\omega_0^{-1}$. 
For the special case $\chi=90\arcdeg$, $\beta>0$,
$\Delta s$ is given exactly by
\begin{equation}
 \Delta s
 =
 \frac{4}{|\epsilon|}
 \left(
  \frac{\beta}{\eta-\gamma}
 \right)^{1/2}
 K\left\{
  \left[
   \frac{\gamma+\eta\beta}{\beta(\eta-\gamma)}
  \right]^{1/2}
 \right\},
 \label{eq:tum27}
\end{equation}
where $K(k)$, a complete elliptic integral of the first kind,
increases logarithmically from $K(0)=\pi/2$ to $K(1)=\infty$.
Note that the precession is not exactly periodic,
because $\Delta s\propto (\eta-\gamma)^{-1/2}$
increases adiabatically on the braking time-scale $\tau_0$.
Indeed,
$\Delta s$ increases significantly and approaches 
$\omega_0\tau_0$
(so that the separation into slow and fast time-scales
in Section \ref{sec:tum2c} breaks down)
under two special sets of circumstances:
(i) at $k=1$, 
where $u_1$, $u_2$ and $u_3$ suddenly swap oscillation
modes (`mode jumping'; see Section \ref{sec:tum3d}); 
and (ii) at the $\eta=\gamma$ resonance,
where the the star rotates steadily with $u_2=u_3=0$
and $u_1={\rm const}$ for $\beta>-1$
(Section \ref{sec:tum3d}).
If the star is triaxial, a second precession time-scale,
$\Delta s \sim |\beta|^{1/2}|\epsilon'|^{-1}$,
is introduced.

\subsection{Precession and Nutation Amplitudes
 \label{sec:tum3b}}
\begin{figure}
\plotone{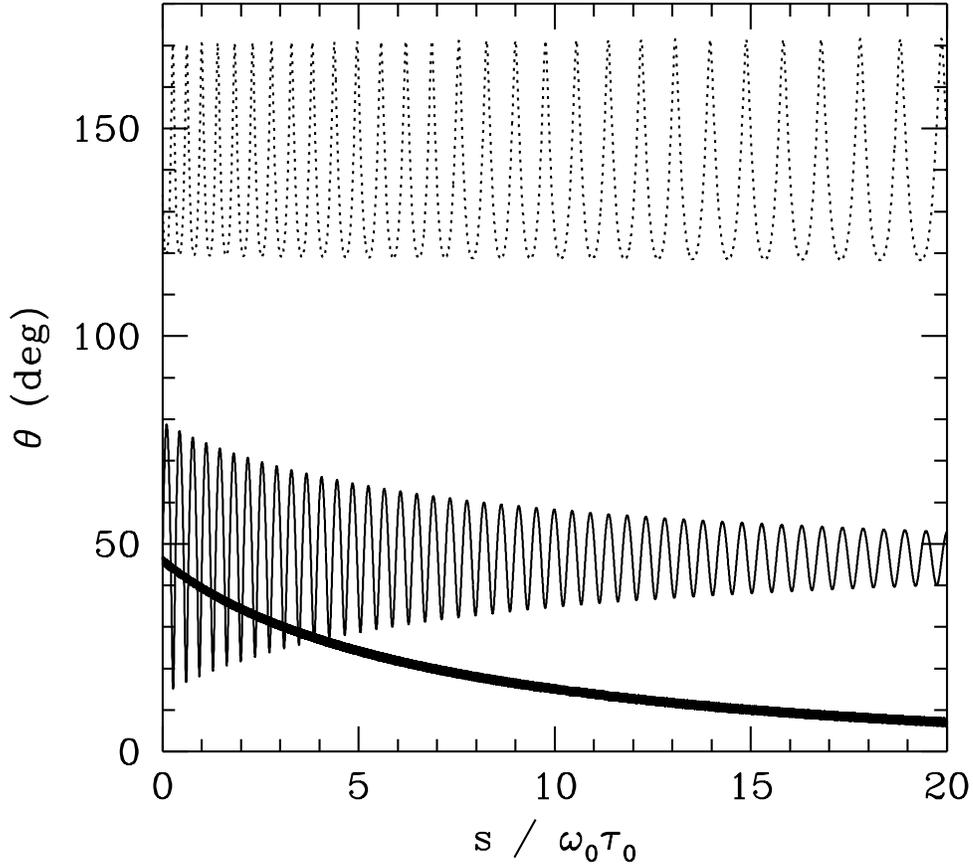}
\caption{
Precession angle $\theta$, in degrees,
as a function of time, in units of the spin-down 
time-scale $\tau_0$.
Thick band:
$\beta=20$, $\chi=20\arcdeg$,
$uG(x_0)=47.4$,
$u_{1,0}=0.387$, $u_{2,0}=0.6$, $u_{3,0}=0.7$.
Although not apparent to the eye,
the band is a rapid oscillation with a peak-to-peak
amplitude of $\approx 1.6\arcdeg$.
Solid curve: 
$\beta=0.8$, $\chi=20\arcdeg$,
$uG(x_0)=47.4$,
$u_{1,0}=0.387$, $u_{2,0}=0.6$, $u_{3,0}=0.7$.
Dotted curve:
$\beta=0.8$, $\chi=20\arcdeg$,
$uG(x_0)=47.4$,
$u_{1,0}=0.387$, $u_{2,0}=0.6$, $u_{3,0}=-0.7$.
}
\label{fig:tum1}
\end{figure}

Figure \ref{fig:tum1} displays the precession angle
$\theta$, defined to be the angle between $\bfw$
and ${\bf e}_3$ ($\cos\theta=u_3/u$), as a function
of time.  We see from Figure \ref{fig:tum1}
that the star precesses ($\theta\neq 0$ on average)
and nutates ($\theta$ oscillates in a range
$\theta_1\leq\theta\leq\theta_2$ during one precession
period), and that the slow evolution of the precession
and nutation amplitudes is determined by $\beta$,
$\chi$ and the initial orientation of $\bfw$.
In the regime $\beta\gg 1$ where the free precession
period $\tau_2$ is shorter than both torque-related
time-scales $\tau_0$ and $\tau_1$,
the nutation amplitude is small and $\theta$
decreases exponentially to zero for $\chi=20\arcdeg$
(thick band in Figure \ref{fig:tum1}).
This result, and a similar calculation for 
$\chi=70\arcdeg$ (not shown), confirm
\markcite{gol70}Goldreich's (1970) conclusion
that the slow evolution of $\theta$
in the regime $\beta\gg 1$
depends solely on $\chi$, with
$\theta\rightarrow 0\arcdeg$ for
$\chi < \chi_{\rm cr}=\cos^{-1}(3^{-1/2})\approx 55\arcdeg$
and $\theta\rightarrow 90\arcdeg$ for
$\chi > \chi_{\rm cr}$.

\markcite{gol70}Goldreich's (1970) conclusion is invalid
when $\tau_1$ and $\tau_2$ are comparable.
The solid and dotted curves in Figure \ref{fig:tum1}
both correspond to $\chi=20\arcdeg < \chi_{\rm cr}$,
but with
$\beta=0.8$ and hence $\tau_1\sim\tau_2$. 
Neither curve behaves as predicted by
\markcite{gol70}Goldreich (1970):
either $\theta$ approaches a constant non-zero value
while the nutation amplitude decreases to zero
(solid curve), or else $\theta$ remains constant on average
with a peak-to-peak nutation amplitude of
$\approx 50\arcdeg$ (dotted curve).
In each case, {\em the precession is persistent},
and its character is determined by the initial orientation
of $\bfw$.

\subsection{Do the Magnetic and Rotation Axes Align?
 \label{sec:tum3c}}
\begin{figure}
\plotone{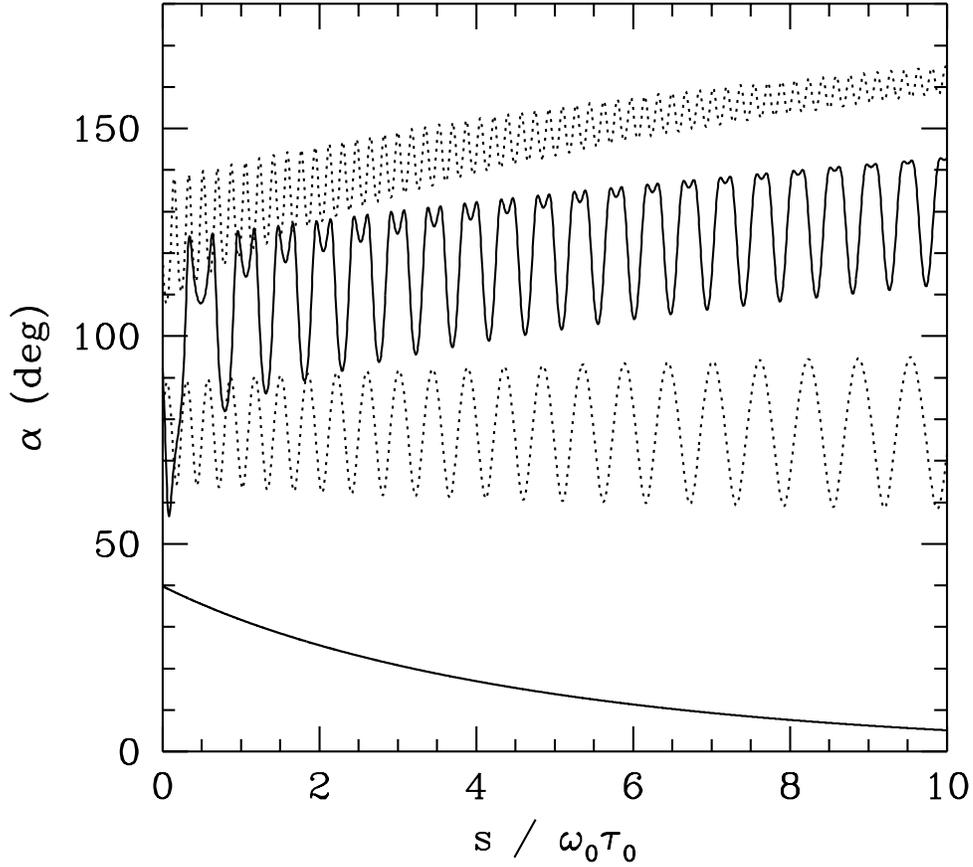}
\caption{
Magnetic inclination angle $\alpha$, in degrees,
as a function of time, in units of the spin-down 
time-scale $\tau_0$, for
$uG(x_0)=47.4$ and initial conditions
$u_{1,0}=-0.5$, $u_{2,0}=0.4$, $u_{3,0}=0.768$.
Lower solid curve:
$\beta=0.5$, $\chi=0\arcdeg$.
Lower dotted curve:
$\beta=0.5$, $\chi=45\arcdeg$.
Upper solid curve:
$\beta=-0.5$, $\chi=60\arcdeg$.
Upper dotted curve:
$\beta=-0.5$, $\chi=90\arcdeg$.
}
\label{fig:tum2}
\end{figure}

The magnetic inclination angle $\alpha$ between $\bfw$
and ${\bf m}$ is defined in terms of $u_1$,
$u_2$ and $u_3$ by
$u\cos\alpha= u_1\sin\chi + u_3\cos\chi$.
In Figure \ref{fig:tum2}, we plot $\alpha$
as a function of time
for several choices of $\beta$ and $\chi$.
For $\chi=0\arcdeg$, $\beta=0.5$, we see that 
$\alpha$ decreases exponentially to zero on the 
braking time-scale. 
By solving (\ref{eq:tum2}) to (\ref{eq:tum4})
analytically for $\chi=0\arcdeg$, one can show that
$\alpha$ approaches zero for arbitrary $\beta$,
implying that a star subject to a predominantly 
magnetic deformation ultimately
becomes an aligned rotator ---
a state in which it cannot be detected as a pulsar. 
An aligned final state ($\alpha=180\arcdeg$) can also be
attained by a star with $\chi\neq0\arcdeg$
(upper dotted curve in Figure \ref{fig:tum2}).
In contrast, if the star nutates persistently
as discussed in Section \ref{sec:tum3b},
$\alpha$ mimics $\theta$ and oscillates within a range
(typically tens of degrees).

\subsection{Oblate and Prolate Stars,
 and the $\eta$-$\gamma$ Phase Plane
 \label{sec:tum3d}}
\begin{figure}
\vspace{2.3cm} \centerline{\hspace{-3.5cm}(a)\hspace{7.6cm}(b)}
 \vspace{-2.3cm} \centerline{\psfig{figure=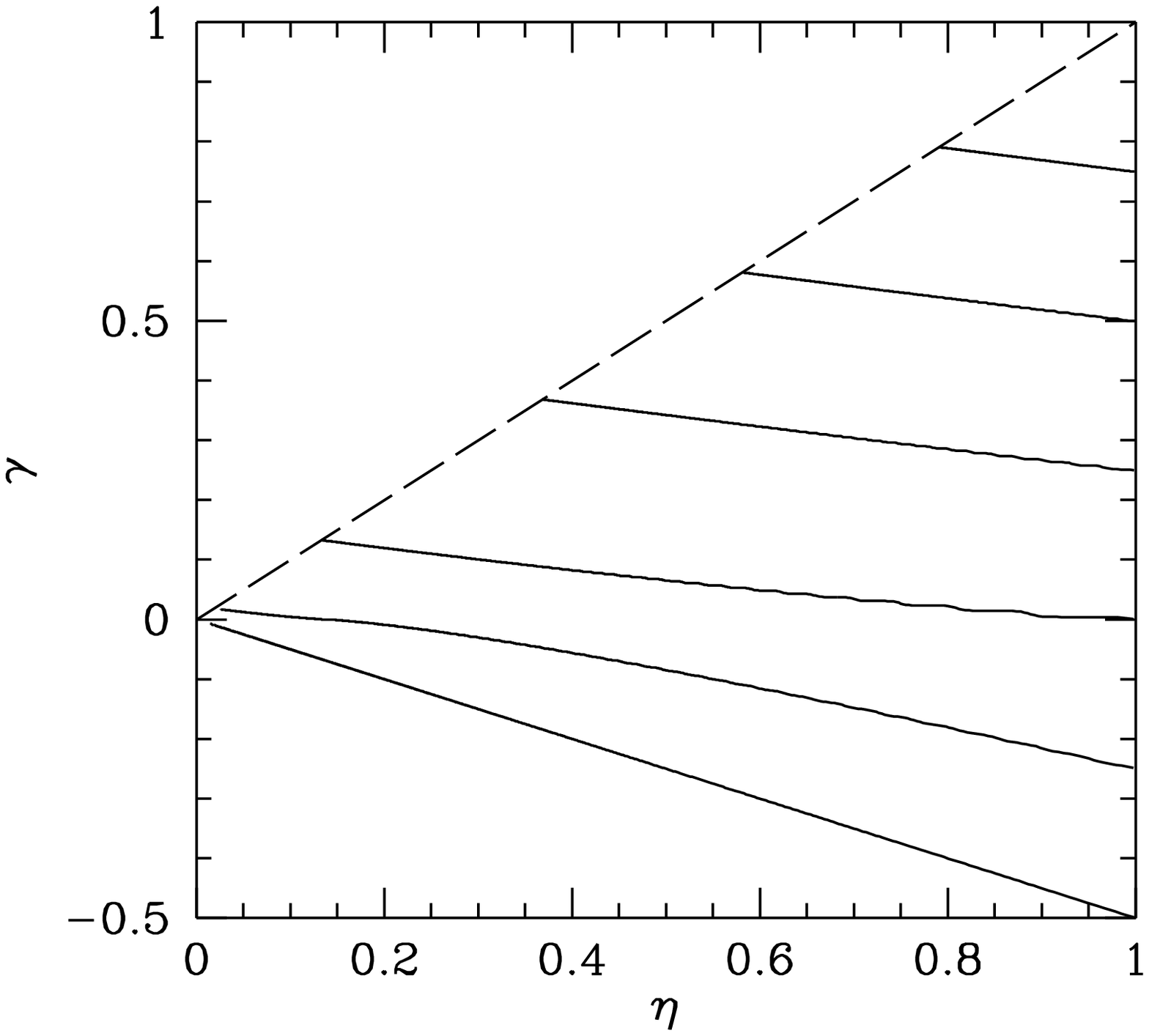,height=8cm}
 \psfig{figure=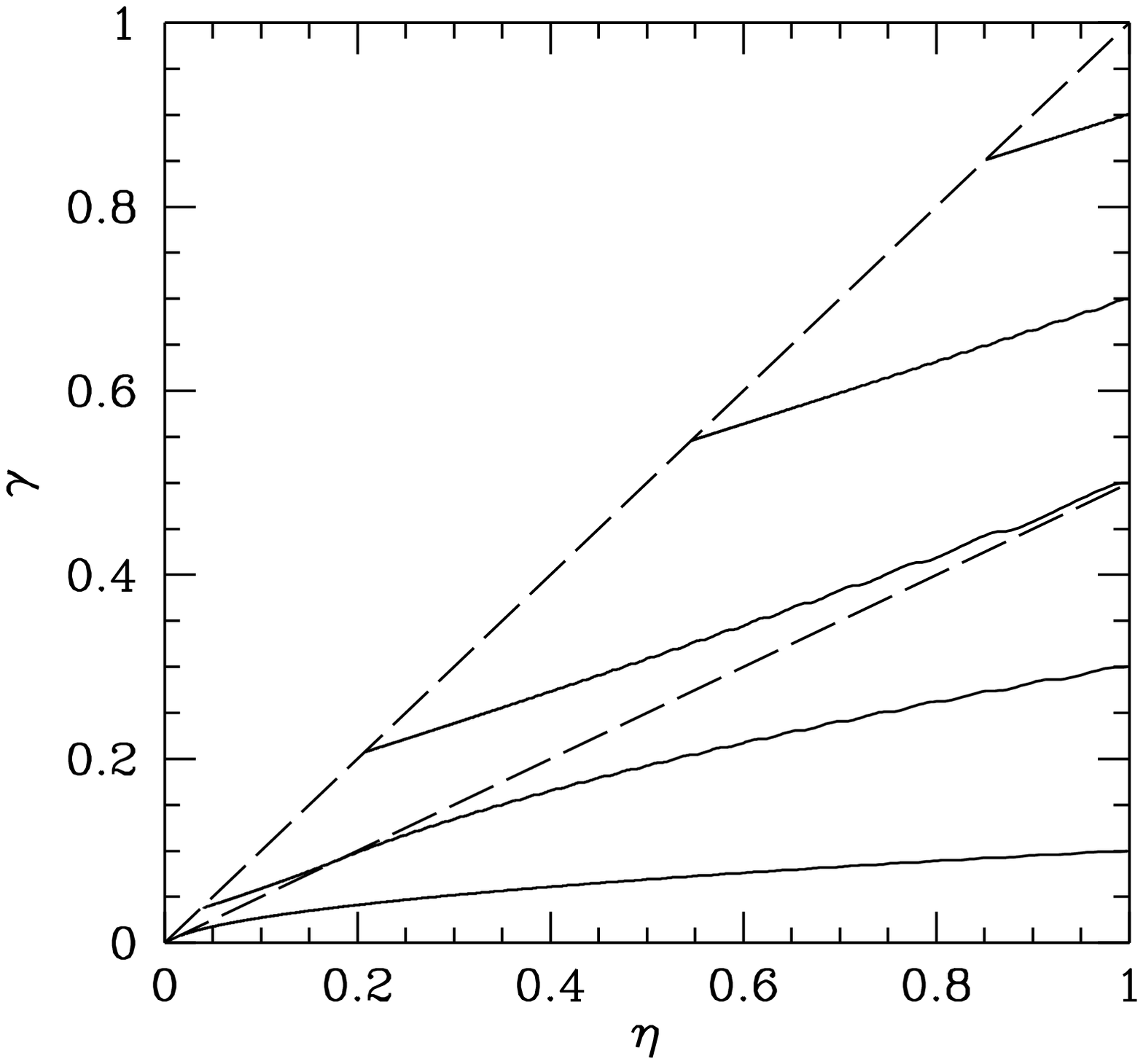,height=8cm}}
\vspace{2.3cm} \centerline{\hspace{-3.5cm}(c)}
 \vspace{-2.3cm} \centerline{\psfig{figure=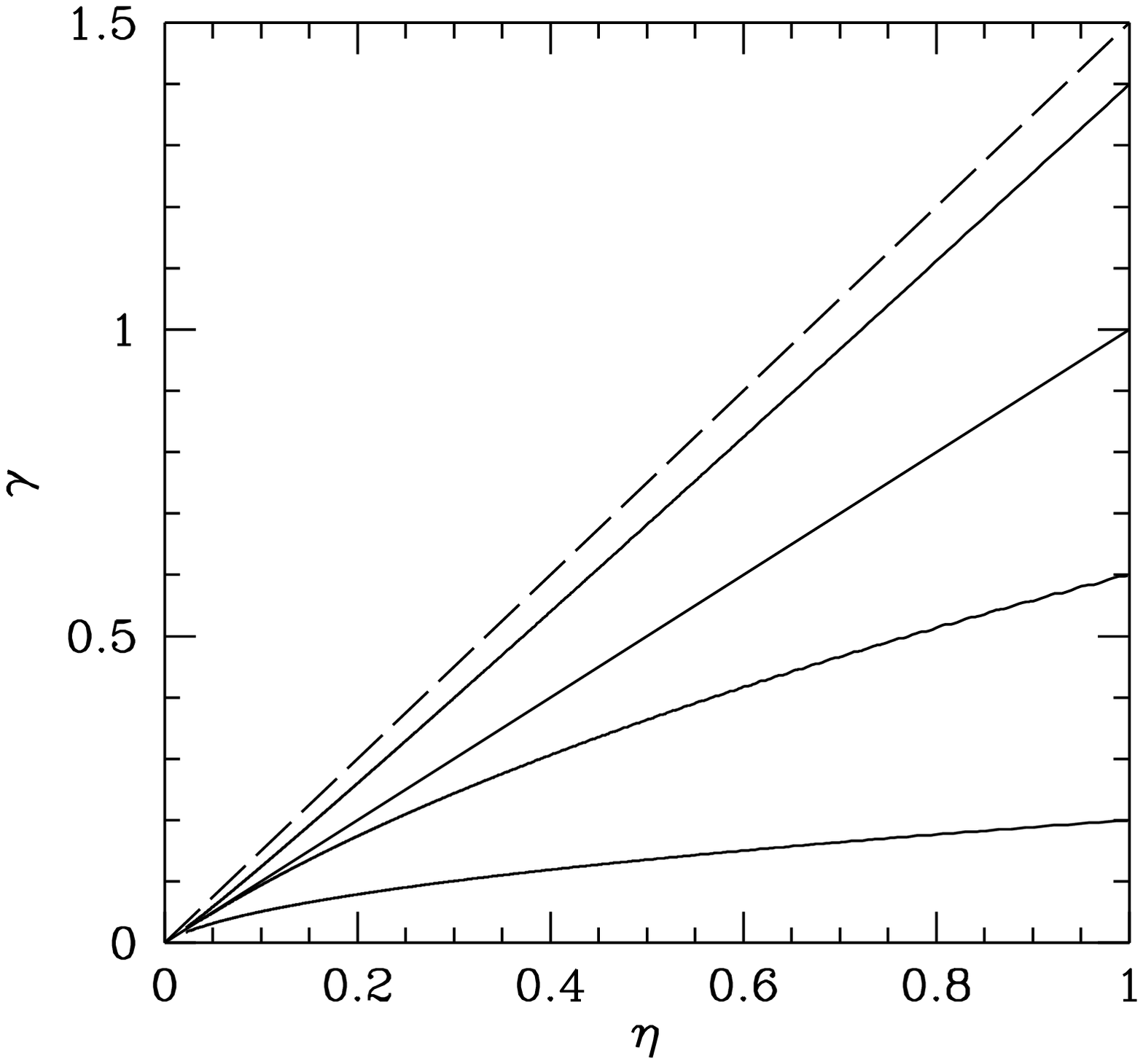,height=8cm}}
\caption{
Phase diagrams showing trajectories in the
$\eta$-$\gamma$ plane, 
for $uG(x_0)=47.4$, $\chi=90\arcdeg$,
and $0\leq s\leq 10^2\omega_0\tau_0$.
The trajectories begin at $\eta=1$,
with different initial values of $\gamma$ corresponding 
to different initial conditions, 
and move from right to left across the diagrams.
The broken diagonals are the lines $\eta =\gamma$
and $\eta=-\beta\gamma$.
Phase diagrams are shown for three stars: 
(a) $\beta=0.5$ (oblate),
(b) $\beta=-0.5$ (prolate),
(c) $\beta=-1.5$ (prolate).
}
\label{fig:tum3}
\end{figure}

An instructive way to view the evolution of the
rotation is to follow the star's
trajectory $[\eta(s),\gamma(s)]$ on the $\eta$-$\gamma$
phase plane. 
As shown in Section \ref{sec:tum2c}, $\eta$
and $\gamma$ are approximately constant 
over one precession period, varying slowly
on the braking time-scale $\tau_0$.
Figure \ref{fig:tum3} shows phase diagrams 
for three stars with $\chi=90\arcdeg$ and
different ellipticities.
Since $\eta$ and
$\gamma$ are not exactly constant over one precession
period, the trajectories are slightly irregular.

Figure \ref{fig:tum3}(a) shows that an {\em oblate} star
($\beta > 0$)
evolves asymptotically towards a state with 
$\gamma=\eta\neq 0$ 
and hence $u_2=u_3=0$, $u_1={\rm const}$
(see [\ref{eq:tum22}] and [\ref{eq:tum23}]);
in other words, for $\chi=90\arcdeg$,
an oblate star always approaches steady-state rotation
with $\bfw$ parallel to ${\bf m}$ ($={\bf e}_1$).
Equations (\ref{eq:tum2}) to (\ref{eq:tum4}) imply
that this is a singular fixed point which exists
for $\chi=90\arcdeg$ only.
The evolution of a {\em prolate} star
($\beta<0$) depends on
the relative magnitudes of the precession periods $\tau_1$ 
(radiative) and $\tau_2$ (free).
When the radiative precession is faster ($-1<\beta < 0$),
as in Figure \ref{fig:tum3}(b),
the star evolves to a state with $\gamma=\eta\neq 0$,
as discussed above.
When the free precession is faster ($\beta < -1$),
as in Figure \ref{fig:tum3}(c),
the star evolves to a state with $\gamma=\eta=0$,
and $\bfw$ does not necessarily align with any
preferred axis on the way, 
although it may do so for specific initial conditions.
In all the above cases, the phase-plane trajectories
are confined within the triangular region 
$0\leq\eta\leq1$, 
${\rm min}(-\eta\beta,0)\leq\gamma 
 \leq {\rm max}(\eta,-\eta\beta)$.

How are the phase diagrams modified for $\chi\neq90\arcdeg$?
Firstly, the trajectories are confined to a smaller
(larger) triangular region for $\beta>0$ ($\beta<0$)
defined by
\begin{equation}
 0\leq\eta\leq 1,
 \phantom{XXX}
 {\rm min}(0,\gamma_1,\gamma_3) \leq\gamma\leq
 {\rm max}(\gamma_1,\gamma_2,\gamma_3),
 \label{eq:tum27a}
\end{equation}
with $\gamma_1=\eta(\cos^2\chi-\beta)$,
$\gamma_2=\eta\sin^2\chi$ and
\begin{equation}
 \gamma_3
 =
 \eta [ \sin^2(\chi+\psi) - \beta\sin^2\psi ],
 \phantom{XXX}
 \tan 2\psi =
 \frac{\sin 2\chi}{\beta - \cos 2\chi}.
 \label{eq:tum27b}
\end{equation}
Secondly, although the $\beta>-1$ trajectories approach
the diagonal line 
$\gamma={\rm max}(\gamma_1,\gamma_2,\gamma_3)$,
they do not stop there like in Figure \ref{fig:tum3}.
Instead, they bend downwards to merge with the diagonal,
travelling down along it as $\eta$ decreases.
This is because the state
$\gamma={\rm max}(\gamma_1,\gamma_2,\gamma_3)$
is only a fixed point for $\chi=90\arcdeg$,
as pointed out above.

\begin{figure}
\plotone{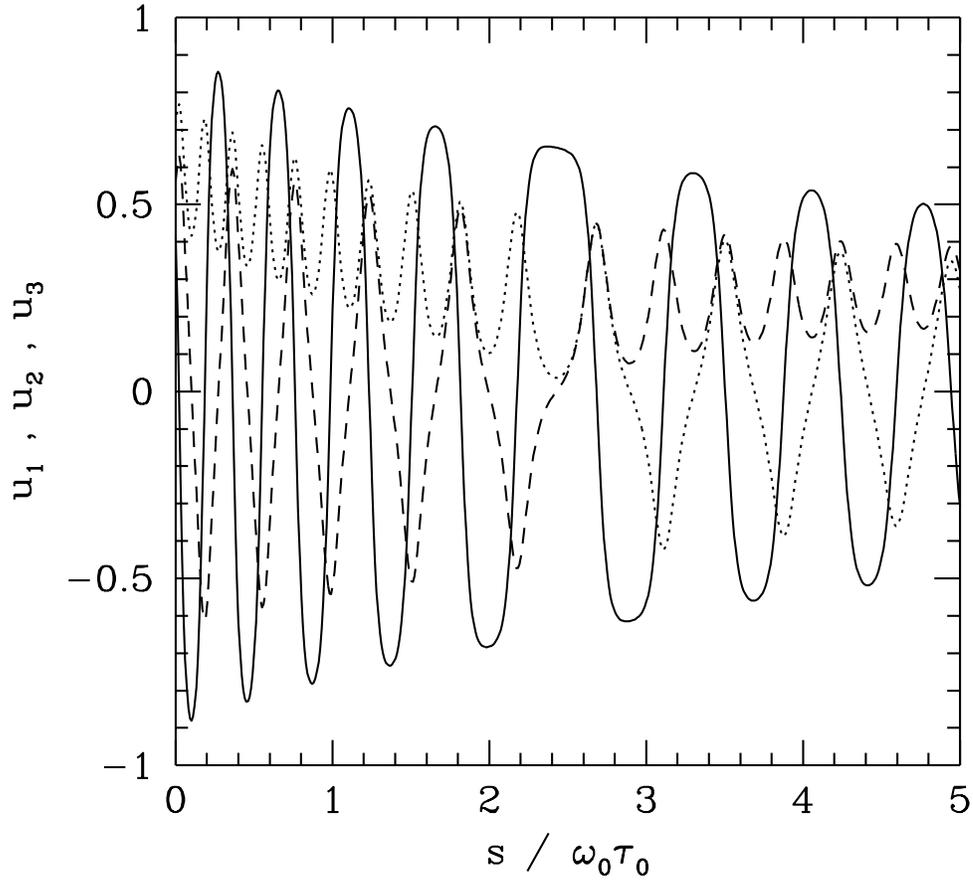}
\caption{
Principal angular frequency components $u_1$ (dashed curve),
$u_2$ (dotted curve) and $u_3$ (solid curve),
in units of $\omega_0$,
as functions of time, in units of the spin-down 
time-scale $\tau_0$. 
Note that
$u_1$ and $u_2$ interchange oscillation modes at
$s\approx2.7\,\omega_0\tau_0$.
All curves are for
$\chi=90\arcdeg$, $\beta=-0.5$, 
$uG(x_0)=47.4$,
$u_{1,0}=0.539$, $u_{2,0}=0.7$, $u_{3,0}=0.469$.
}
\label{fig:tum4}
\end{figure}

The trajectories in Figures \ref{fig:tum3}(a)--(c)
asymptotically approach, or travel exactly along,
the lines $\gamma=\eta$ and $\gamma=-\eta\beta$
except in the case $-1<\beta<0$
(Figure \ref{fig:tum3}[b]), 
where trajectories with $\gamma<-\beta$ at $s=0$
subsequently cross the line $\gamma=-\eta\beta$.
When this happens, a phenomenon we call
`mode jumping' takes place.
As shown in Figure \ref{fig:tum4}, $u_1$ and $u_2$ 
interchange oscillation modes at
$s\approx2.7\,\omega_0\tau_0$, 
swapping between a wine-glass mode and a sinusoidal mode,
while the $u_3$ oscillation becomes
temporarily flatter-peaked.

\subsection{Pseudo-Glitches in the Frequency Derivative
 of a Triaxial Star
 \label{sec:tum3e}}
\begin{figure}
\plotone{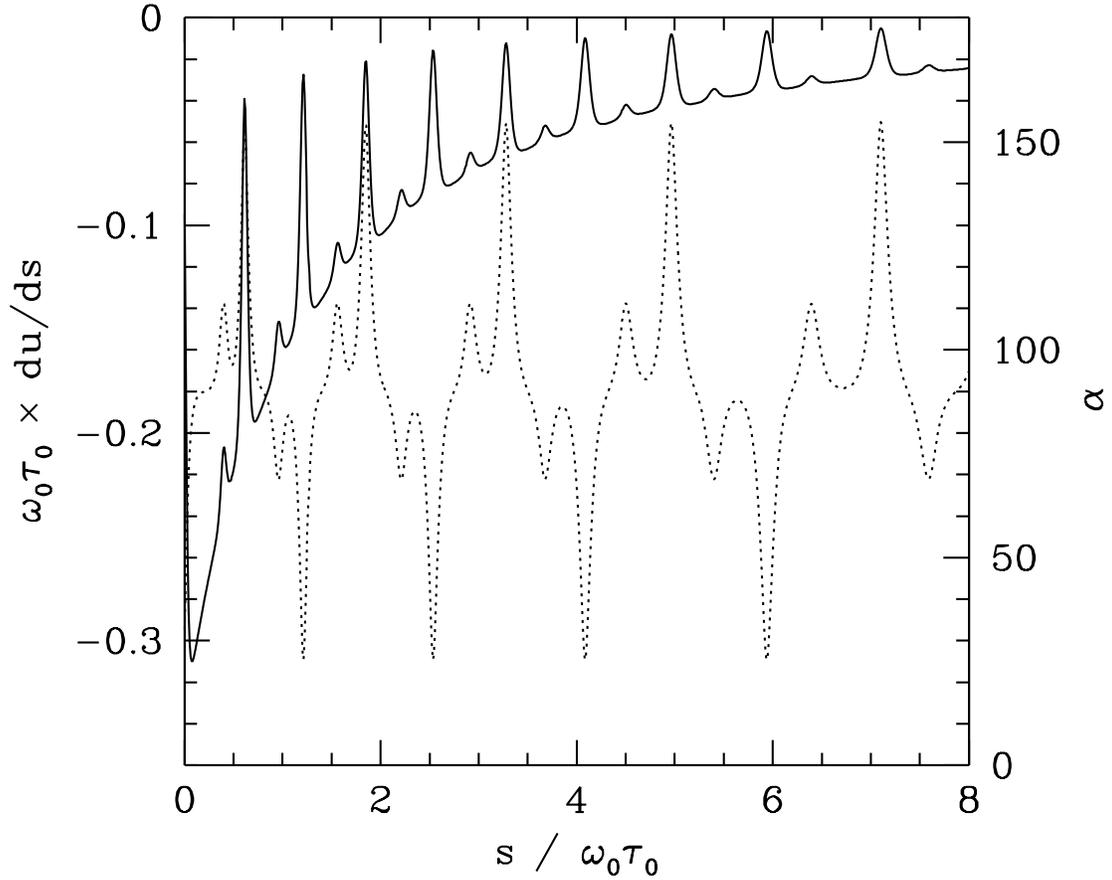}
\caption{
Angular frequency derivative $\dot{u}$ (solid curve), 
in units of $(\omega_0\tau_0)^{-1}$,
and magnetic inclination angle $\alpha$ (dotted curve),
in degrees,
as functions of time, 
in units of the braking time-scale $\tau_0$.
The local minima and maxima of $\dot{u}$ and $\alpha$
correspond one-to-one.
Both curves are for
$\chi=40\arcdeg$,
$\beta=9.5$, $\epsilon'=0.09\epsilon$,
$uG(x_0)=9.8$,
$u_{1,0}=0.873$, $u_{2,0}=0.28$, $u_{3,0}=0.4$.
}
\label{fig:tum5}
\end{figure}

A triaxial ellipsoid of inertia arises naturally  if,
for example, the deformation is predominantly magnetic 
with appreciable quadrupolar and off-centred components,
or if one has $\epsilon^{\rm cr}\sim\epsilon^{\rm mag}$.
Figure \ref{fig:tum5} plots the angular frequency derivative
$\dot{u}$ as a function of time for a triaxial star with
$\epsilon'=0.09\epsilon$.
We see that the smooth, braking-related decrease
of $|\dot{u}|$ is punctuated by
sudden, quasi-periodic spikes in which $|\dot{u}|$ 
changes by up to 90 per cent.
We call these excursions `pseudo-glitches'.
They resemble true glitches because
(i) they recur quasi-periodically with period
$\sim x_0 \tau_0$ 
($\approx 10\,{\rm yr}$ for the Crab),
and (ii) $\dot{u}$ returns to its trend value after each
excursion.
However, they are manifestly not true glitches because
(i) their rise time is too long 
(cf.\ $\lesssim 10^{-10}\tau_0$ for Crab glitches),
and (ii) they do not cause $u$ itself to increase.
Note that $\dot{u}$ oscillates about the average
spin-down trend even for $\epsilon'=0$, but not in the
spiky fashion of Figure \ref{fig:tum5}.

What is the physical origin of pseudo-glitches?
The dotted curve in Figure \ref{fig:tum5} shows that
pseudo-glitches coincide with rapid changes in $\alpha$,
accompanied by mode jumping,
which occur when the free precession
is modulated on a fast time-scale
$(\epsilon'/\epsilon)\tau_2$
which couples resonantly to the radiative precession
(i.e.\ $\epsilon\tau_1\approx\epsilon'\tau_2$).
It turns out that the sharpness of the spikes is sensitive
to $\epsilon'$ and $uG(x_0)$;
for the example in Figure \ref{fig:tum5}, the spikes are
washed out once $\epsilon'$ falls outside the range
$0.05\lesssim \epsilon'/\epsilon \lesssim 0.15$.
Figure \ref{fig:tum5} shows the case $\chi=40\arcdeg$.
For larger $\chi$, the small bumps between the
spikes increase in amplitude until they become spiky
themselves. For smaller $\chi$, the bumps
flatten until they disappear.

\subsection{Corotating Magnetosphere
 \label{sec:tum3f}}
The foregoing results pertain to the regime $x_0\ll 1$,
where $r_0$ is taken to be the stellar radius $R$.
However, recent work by
\markcite{mel97}Melatos (1997) suggests
that the corotating magnetosphere of a neutron star 
acts as a perfectly conducting, rigid extension of the
stellar interior out
to a characteristic radius $r_{\rm v}$
where outflowing plasma is not
constrained to flow along magnetic field lines by
cyclotron losses, and that
it is therefore necessary to set $r_0=r_{\rm v}$ when
calculating electromagnetic spin-down properties 
like pulsar braking indices.
For young pulsars (e.g.\ the Crab, PSR B1509$-$58, PSR B0540$-$69), 
one finds $r_{\rm v}\lesssim c/\omega$, and hence 
$x_0\lesssim 1$.

In the regime $x_0\lesssim 1$, the radiative precession
period satisfies $\tau_1\sim\tau_0$,
the near-field radiation torque uncouples from the free
precession, and the evolution of the precession
amplitude is governed completely by $\chi/\chi_{\rm cr}$
(\markcite{gol70}Goldreich 1970).
However, a pulsar born with $x_0\lesssim 1$
soon evolves towards the regime $x_0\ll 1$
as $\omega$ decreases.
Therefore, unless the free precession amplitude
decreases to zero during the initial braking phase
with $x_0\lesssim 1$ ({\em not} the
outcome in general), the rotation behaves thereafter in the way
described in Sections \ref{sec:tum3a} to \ref{sec:tum3e}.

\section{APPLICATION TO OBSERVATIONS 
 \label{sec:tum4}}
Timing and polarization studies
suggest that most, if not all,
young pulsars do not precess in the manner described in
Sections \ref{sec:tum2} and \ref{sec:tum3}.
The implication is that isolated neutron stars 
born with a large precessional motion
approach stable spin states ($\bfw$ parallel to a
principal axis of inertia) over times that are short compared 
to their current ages, probably due to internal friction.
We outline the conditions for a neutron star to be born
with a large precessional motion in Section \ref{sec:tum4a1}, 
summarize the status of observational searches for
precession in Section \ref{sec:tum4a}, 
and discuss the observational consequences of
frictional stabilization
in Section \ref{sec:tum4b}.

\subsection{Precession at Birth
 \label{sec:tum4a1}}
When a neutron star is born, it spins about an axis $\bfw_0$
dictated by conservation of angular momentum (of the degenerate
remnant and the ejecta) during the supernova explosion.
There is no reason why $\bfw_0$ should immediately be parallel
to the principal axis of greatest non-hydrostatic moment
of inertia (the magnetic axis ${\bf m}$, since the star is
a fluid). Indeed, \markcite{tho93}Thompson \& Duncan (1993)
argued that post-collapse convection destroys any correlation
between $\bfw_0$ and ${\bf m}$.

Viscous dissipation in the fluid star
forces $\bfw_0$ to approach ${\bf m}$ over time.
The dissipation rate is therefore critical in determining
whether the star will exhibit a large initial precession
when its crust crystallizes.
Unfortunately, the viscosity of a newly born neutron star 
is poorly constrained. 
\markcite{cut87}Cutler \& Lindblom (1987) estimate
the viscous damping time (e.g.\ for stellar oscillations)
to be roughly
$3\times 10^{2} 
 (\rho/10^{17}\,{\rm kg\,m^{-3}})^{-1.25}
 (T/10^9\,{\rm K})^{2}
 {\rm yr}$,
to be compared with a crust crystallization time of less
than one year,
but this estimate is known to be valid only in narrow ranges of 
density $\rho$ and temperature $T$ centred on the above fiducial
values.
In what follows, we make no assumption about the viscous
damping time and instead explore several possible scenarios.

If the viscosity is high enough, $\bfw_0$ aligns with ${\bf m}$
first, before the crust crystallizes. Assuming that the symmetry axis
of the crust when it crystallizes is along $\bfw_0$
(likely, though not certain), then the principal axis ${\bf e}_3$
(from both elastic and magnetic contributions) is parallel
to $\bfw_0$, and there is no precession.

If the viscosity is low enough, the crust crystallizes first,
before $\bfw_0$ has time to align with ${\bf m}$ ---
the order of events implied by the viscosity estimates of
\markcite{cut87}Cutler \& Lindblom (1987). 
In this scenario, two things can happen:
(i) one has $\epsilon^{\rm cr} \gg \epsilon^{\rm mag}$,
and the symmetry axis of the crust when it crystallizes
is along $\bfw_0$,
so that ${\bf e}_3$ is parallel to $\bfw_0$ and there is no
precession;
or (ii) one has $\epsilon^{\rm cr} \ll \epsilon^{\rm mag}$,
so that ${\bf e}_3$ is parallel to ${\bf m}$ 
--- which is {\em not} parallel to $\bfw_0$ at the epoch of
crystallization ---
and there is a large precession.
Below we explore the observational implications of
scenario (ii).

\subsection{Changes in Pulse Profile and Polarization
 \label{sec:tum4a}}
The results of Section \ref{sec:tum3} imply that a
neutron star in an arbitrary initial spin state
precesses (and nutates) with period $\sim x_0\tau_0$
and a typical amplitude of tens of degrees,
and that the motion is persistent in general.
One therefore expects fractional changes of order unity
in the pulse profile (e.g.\ relative height or separation
of conal components),
magnetic inclination angle $\alpha$
(measured from polarization-swing data),
and angular frequency derivative $\dot{\omega}$
(measured by timing) over
a single precession period.
For young, Crab-like objects
($x_0\approx 10^{-2}$, $\tau_0\approx 10^3\,{\rm yr}$),
the fractional changes amount to
$\sim 10$ per cent per year and ought to be readily observable;
for old objects
($x_0\approx 10^{-3}$, $\tau_0\approx 10^6\,{\rm yr}$),
the changes amount to $\sim 0.1$ per cent per year
and are harder to detect.

Contrary to the above expectation,
the only reliable detection of neutron-star precession 
to date has been
\markcite{wei89}Weisberg et al.'s (1989)
discovery of the general relativistic
geodetic precession of PSR 1913$+$16.
Six years of accurate measurements of the doubly-peaked
radio pulse profile
revealed that the flux density in the first component
is decreasing relative to
the second by $\approx 1$ per cent per year,
consistent with the line of
sight moving across a spot of enhanced emissivity
in the magnetosphere at the rate prescribed by geodetic
precession.
\markcite{wei89}Weisberg et al.'s (1989)
upper limit $|\Delta w| < 0\arcdeg.06$
on the six-year change in pulse width $w$
implies a maximum precession amplitude
$\lesssim 0\arcdeg.4\,{\rm yr^{-1}}$,
consistent with a relativistic origin
and much larger than the predicted radiative precession
amplitude of $\approx 0\arcdeg.001\,{\rm yr^{-1}}$.
Hence PSR B1913$+$16 does not usefully constrain
the amplitude of radiative precession in old pulsars.

No unambiguous instances of non-relativistic precession
are known.
\markcite{lyn88b}Lyne et al.\ (1988) reported a
quasi-sinusoidal variation 
with a period of $\approx 20$ months in the phase
residuals of six years of Crab timing data,
and a similar feature was claimed to exist in Vela
(\markcite{mcc90}McCulloch et al.\ 1990).
However, the variation may be an artifact of an
unexpectedly high $\ddot{\omega}$ during the
exponential relaxation following an overlooked glitch.
The quasi-periodic nature of the Vela glitches
(\markcite{lyn96}Lyne et al.\ 1996)
is also suggestive of precession,
but the observed $\dot{\omega}$
as a function of time
does not resemble Figure \ref{fig:tum5}.
\markcite{ulm94}Ulmer (1994) reported that the intensity
ratio of the two peaks in the gamma-ray pulse of the
Crab seems to vary sinusoidally with a period
of $\approx 14\,{\rm yr}$ in both
the 50--500\,keV and 50\,MeV bands
(but not in the optical).
However, the 14-yr period differs from that observed by
\markcite{lyn88b}Lyne et al.\ (1988)
and a 60\,s modulation of the Crab's optical
pulses found by
\markcite{cad97}\v{C}ade\v{z} et al.\ (1997).

Direct measurements of $\alpha$ from pulsar polarization
swings do not show any evidence for secular changes
on a yearly time-scale. It has been claimed
that changes in
$\alpha$ on a time-scale of $\sim 10^4\,{\rm yr}$
explain the braking indices of the Crab and Vela,
and sudden jumps in $\alpha$ 
explain the persistent increase in $\dot{\omega}$
following a glitch
(\markcite{all97}Allen \& Horvath 1997;
\markcite{lin97}Link \& Epstein 1997),
but the relevant time-scales are too long and too short,
respectively, to be precession-related.
\markcite{tau97}Tauris \& Manchester (1997) 
used polarization-swing data for more than 100 pulsars
to construct the $\alpha$ distribution of the
pulsar population, corrected for 
decreasing beam radius with age.
The corrected distribution is skewed towards
small $\alpha$, with $\langle\alpha\rangle$ decreasing
on a time-scale of $\approx 10^7\,{\rm yr}$ ---
once again, too gradual to be a precession-related effect.

The observational evidence against substantial changes
in pulse profiles and polarization properties implies that
any precession initially present is damped rapidly.
Consequently, direct detection of a precessing, isolated
pulsar may only be possible in the immediate
aftermath of a Galactic supernova.
If the precession amplitude is sufficiently large,
the newly born pulsar will shine intermittently, as
its emission cone drifts into and out of the line of
sight, as well as exhibit the profile and polarization
changes discussed above.

\subsection{Crustal Heating by Internal Friction 
 \label{sec:tum4b}}
Frictional damping inside a neutron star proceeds rapidly
in theory. Several authors have estimated the
dissipation times due to time-dependent elastic strain
(\markcite{cha71}Chau \& Henriksen 1971;
\markcite{mac74}Macy 1974)
and imperfect core-crust coupling 
(\markcite{sha77}Shaham 1977;
\markcite{dec80}de Campli 1980;
\markcite{alp87}Alpar \& \"{O}gelman 1987;
\markcite{lin93}Link et al.\ 1993)
and found them to be small compared to the present ages
of young, Crab-like pulsars.
(The estimates assume a small precession amplitude;
cf.\ Section \ref{sec:tum3}.)
In addition, there is the analogy of friction inside
the Earth. Dynamic satellite measurements have revealed
that the Earth's non-hydrostatic ellipsoid of inertia
has its ${\bf e}_3$ axis parallel to $\bfw$ to an
excellent approximation\footnote{I thank P.\ Goldreich
for bringing this fact to my attention.}
(\markcite{lam80}Lambeck 1980, p.\ 31),
and that the direction of ${\bf e}_3$ fluctuates by
less than $1\arcsec$ under the action of solar and
lunar tides, compared to $\approx 10\arcdeg$
for the ${\bf e}_1$ and ${\bf e}_2$ axes
(\markcite{bur93}Bur\v{s}a \& P\v{e}\v{c} 1993, p.\ 227).
Clearly, any drift of $\bfw$ away from ${\bf e}_3$
is rapidly damped.

In this paper, we are not concerned with the 
precise origin of the friction in a neutron star;
we simply suppose it exists and examine the fate of the
dissipated energy. Initially, when the star is precessing,
its angular momentum and energy are given by
$L_{\rm i}=
 [ I_1^2(\omega_{1{\rm i}}^2+\omega_{2{\rm i}}^2)
 + I_3^2 \omega_{3{\rm i}}^2 ]^{1/2}$
and
$E_{\rm i}=
 \case{1}{2}
 I_1(\omega_{1{\rm i}}^2+\omega_{2{\rm i}}^2)
 + \case{1}{2}
 I_3\omega_{3{\rm i}}^2$.
After the precession has been damped,
the final angular momentum and energy 
are given by $L_{\rm f}=I_3\omega_{3{\rm f}}$
and $E_{\rm f}=\case{1}{2}I_3\omega_{3{\rm f}}^2$
($\bfw\rightarrow{\bf e}_3$ by analogy with the Earth). 
If the damping occurs over a time $\tau_{\rm d}$
that is short compared to $\tau_0$ and $\tau_1$,
we have $L_{\rm i}\approx L_{\rm f}$ and hence
\begin{equation}
 \Delta E =
 -\case{1}{2} \epsilon I_1
 (\omega_{1{\rm i}}^2+\omega_{2{\rm i}}^2).
 \label{eq:tum28}
\end{equation}
For $\omega_{1{\rm i}}\sim\omega_{2{\rm i}}\sim\omega_0$
(Section \ref{sec:tum3}),
the total dissipated energy is of order
$\epsilon I_1 \omega_0^2$.

We now suppose that the dissipation is localized
in the inner and outer crusts, where the ions
are organized into a lattice and the shear modulus is
non-zero, and we estimate the resulting increase in the
star's thermal luminosity $L$.
The thermal conduction time in the crust,
$\tau_{\rm cond}$, is not known with certainty
(\markcite{nom87}Nomoto \& Tsuruta 1987),
so we appeal to the extreme cases of slow cooling
($\tau_{\rm d}\ll\tau_{\rm cond}$)
and fast cooling
($\tau_{\rm d}\gg\tau_{\rm cond}$)
to place bounds on $L$.
In the regime $\tau_{\rm d}\ll\tau_{\rm cond}$,
the dissipated energy $\Delta E$ is stored in the
crust as heat for a time $\tau_{\rm cond}$.
The crustal heat capacity at a temperature $T$
is given by
$c_{v}\approx 312 k_{\sc b} ( T / \theta_{\sc d})^3$
(\markcite{sha83}Shapiro \& Teukolsky 1983, p.\ 100)
in the regime where the Debye temperature 
$\theta_{\sc d}\approx 2\times 10^{10}\,{\rm K}$
satisfies $T\ll \theta_{\sc d}$
(i.e.\ at a density 
$\rho\approx 1\times 10^{14}\,{\rm g\,cm^{-3}}$).
Taking $\Delta E=\epsilon I_1 \omega_0^2$,
we find that the final temperature of
an iron crust of mass $10^{-3}M$ is
$T_{\rm f}=
 2\times 10^9 (\epsilon\omega_0^2)^{1/4}\,{\rm K}$,
and its thermal luminosity is
$L=4\pi R^2\sigma T_{\rm f}^4=
 1\times 10^{46}\epsilon\omega_0^2\,{\rm erg\,s^{-1}}$.
In the opposite regime 
$\tau_{\rm d}\gg\tau_{\rm cond}$,
the dissipated heat is conducted rapidly
through the crust
and one has
$L=\Delta E/\tau_{\rm d}=
 1\times 10^{45}\epsilon\omega_0^2
 \tau_{\rm d}^{-1}\,{\rm erg\,s^{-1}}$
(cf.\ \markcite{dec80}de Campli 1980, p.308),
less than the former value for 
$\tau_{\rm d} > 0.1\,{\rm s}$.

The above estimates imply that 
the minimum thermal luminosity of a newly born pulsar
due to frictional damping of its radiative precession is 
\begin{equation}
 L \approx 
 3\times10^{31}
 \left( \frac{\epsilon}{10^{-12}} \right)
 \left( \frac{\omega}{10^3\,{\rm rad\,s^{-1}}} \right)^2 
 \left( \frac{\tau_{\rm d}}{1\,{\rm yr}} \right)^{-1}
 {\rm erg\,s^{-1}}~.
 \label{eq:tum29}
\end{equation}
This ought to be detectable given
$\tau_{\rm d}\lesssim 10^{-3}\,{\rm yr}$
even if there is significant magnetospheric X-ray emission
beamed towards the observer and with
$\tau_{\rm d}\lesssim 3\,{\rm yr}$
if there is not.
The duration of the thermal source is the maximum of 
$\tau_{\rm d}$ and $\tau_{\rm cond}$, starting from
the time when the star first crystallizes into a
body with a rigid crust spinning down electromagnetically.
If the crystallization epoch occurs very shortly after
the supernova explosion itself, it may not be possible
to detect $L$ at all.

\section{CONCLUSION 
 \label{sec:tum5}}
In this paper, the rotation of a rigid, aspherical,
internally frictionless neutron star is analyzed.
We show that, in general, the free precession period 
$\tau_2$ due to elastic and magnetic deformations 
is comparable to the radiative precession period
$\tau_1$ associated with the near-field component
of the radiation torque.
In other words, the `free' precession is not truly
free, a fact that has
been overlooked in the literature to date.

In the regime $\tau_1\sim\tau_2$,
the star rotates in a distinctive way:
(i) it precesses and nutates anharmonically, typically
with an amplitude of tens of degrees (Section \ref{sec:tum3b});
(ii) the magnetic inclination angle $\alpha$ swings through
tens of degrees during one precession period
(Section \ref{sec:tum3c});
(iii) the precession can persist or decay to zero (i.e.\
steady rotation) depending on the parameters $\beta$
and $\chi$ and the initial orientation of $\bfw$
(Section \ref{sec:tum3d});
and (iv) the frequency derivative $\dot{\omega}<0$
oscillates about its overall spin-down trend,
exhibiting spiky, glitch-like behaviour for triaxial 
stars with
$\epsilon\tau_1\approx\epsilon'\tau_2$
(Section \ref{sec:tum3e}).

The precession and nutation lead to fractional changes
of order unity in the pulse profile, polarization swing
and $\dot{\omega}$ of an isolated pulsar on a
time-scale $\sim x_0\tau_0\ll \tau_0$, with
$x_0\tau_0\approx 10\,{\rm yr}$ for young, Crab-like
objects and
$x_0\tau_0\approx 10^3\,{\rm yr}$ for old pulsars.
Such changes are not observed. 
One plausible explanation is that a young neutron star
has $\bfw$ parallel to ${\bf e}_3$ when its crust
crystallizes shortly after birth --- but this is not true
for neutron stars with large hydromagnetic deformations, 
given current viscosity estimates
(Section \ref{sec:tum4a1}).
Another
explanation is that the precession and nutation are
damped by internal friction, perhaps due to
time-dependent elastic strains in the crust.
If the damping takes place over a time $\tau_{\rm d}$,
we show (Section \ref{sec:tum4b}) that the dissipated
energy $\Delta E\approx \epsilon I_1 \omega_0^2$
either heats the crust to a temperature
$T_{\rm f}=
 2\times 10^9 (\epsilon\omega_0^2)^{1/4}\,{\rm K}$
for $\tau_{\rm d}\ll \tau_{\rm cond}$,
yielding a thermal X-ray luminosity
$L=
 1\times 10^{46}\epsilon\omega_0^2\,{\rm erg\,s^{-1}}$,
or else is conducted rapidly to the surface,
yielding
$L=
 1\times 10^{45}\epsilon\omega_0^2
 \tau_{\rm d}^{-1}\,{\rm erg\,s^{-1}}$.
The luminosity $L$  may be detectable
depending on how soon after the supernova explosion the
neutron-star crust crystallizes, and the intensity
of magnetospheric X-ray emission at that epoch.


\acknowledgments
I thank Sterl Phinney, Peter Goldreich and Jon Arons
for comments.
This work was supported by NASA Grant NAG5--2756,
NSF Grant AST--93--15455,
and the Miller Institute for Basic Research in Science.



\clearpage
\appendix
\section{DERIVATION OF EULER'S EQUATIONS 
 \label{sec:tumappa}}
We ignore the slight distortion of the star from a perfectly
spherical figure when calculating its radiation fields and
the radiation-reaction torque.
The electromagnetic fields 
${\bf E}({\bf x},t)$ and ${\bf B}({\bf x},t)$
generated by a magnetized, conducting sphere 
rotating {\em in vacuo} were derived using a multipole
method by \markcite{deu55}Deutsch (1955)
and subsequently corrected for minor typographical errors by
\markcite{mel97}Melatos (1997);
see also \markcite{kab81}Kaburaki (1981).
The radiation torque exerted on the rotating sphere
can be calculated from 
${\bf E}({\bf x},t)$ and ${\bf B}({\bf x},t)$
by integrating the angular momentum flux vector over any
surface $S$ enclosing the sphere:
\begin{equation}
 {\bf N}
 =
 \varepsilon_0 \int_S \left[ 
  ({\bf x}\times{\bf E})\,{\bf E}\cdot d{\bf S} +
  c^2 ({\bf x}\times{\bf B})\,{\bf B}\cdot d{\bf S} -
  \case{1}{2}(E^2 + c^2 B^2)({\bf x}\times d{\bf S})
 \right]~.
 \label{eq:tumapp1}
\end{equation}
Let the radius of the sphere be $r_0$, let its angular
frequency be $\omega$, let $\alpha$
denote the angle between its rotation and
magnetic axes, and assume that
the frozen-in magnetic field 
is dipolar, with polar magnitude $B_0$.
Then, in a Cartesian coordinate system $(x,y,z)$
where the $z$ axis is oriented along the instantaneous
rotation axis $\bfw$ and the magnetic axis 
${\bf m}$ simultaneously 
lies in the $x$-$z$ plane, the instantaneous
radiation torque assumes the form
\begin{equation}
 (N_x,N_y,N_z)
 =
 \frac{2\pi B_0^2 r_0^6 \omega^3} {\mu_0 c^3}
 \left[
  \sin\alpha\cos\alpha F(x_0),
  \sin\alpha\cos\alpha G(x_0),
  -\sin^2\alpha F(x_0)
 \right]~,
 \label{eq:tumapp2}
\end{equation}
with
\begin{equation}
 F(x_0)
 =
 \frac{x_0^4}{5(x_0^6 - 3x_0^4 + 36)}
 +
 \frac{1}{3(x_0^2 + 1)},
 \label{eq:tumapp3}
\end{equation}
\begin{equation}
 G(x_0)
 =
 \frac{3(x_0^2 + 6)}{5x_0 (x_0^6 - 3x_0^4 + 36)}
 +
 \frac{3 - 2x_0^2}{15 x_0 (x_0^2 + 1)}.
 \label{eq:tumapp4}
\end{equation}
Strictly speaking, it is incorrect to refer to
(\ref{eq:tumapp2}) as an instantaneous torque,
because it is calculated under the assumption that the
star has been rotating, and will continue to rotate,
at a constant angular frequency $\omega$; formally,
it is assumed that
the radiation fields exist for all $t$
and are proportional to $e^{i\omega t}$, 
as for an infinitely massive star.
However, the approximation is an excellent one
for a neutron star, where the braking and precession
time-scales $\tau_0$, $\tau_1$, $\tau_2$ 
(Sections \ref{sec:tum2b1}, \ref{sec:tum2b2})
satisfy $\tau_0,\tau_1,\tau_2 \gg \omega^{-1}$.

The form factors (\ref{eq:tumapp3}) and (\ref{eq:tumapp4})
differ from the expressions $F(x_0)=1/3$ and
$G(x_0)=1/2x_0$ appearing in previous works
(\markcite{dav70}Davis \& Goldstein 1970;
\markcite{gol70}Goldreich 1970).
There are two reasons for this difference:
(i) previous authors only included terms of leading
order in the small parameter $x_0$ 
(cf.\ Section \ref{sec:tum3f}), whereas
(\ref{eq:tumapp3}) and (\ref{eq:tumapp4})
are exact for arbitrary $x_0$;
and (ii) previous authors modelled the star as a sphere
of uniform internal magnetization, whereas in this paper
the star is modelled as a perfectly conducting sphere
with a point magnetic dipole at its centre
(\markcite{deu55}Deutsch 1955), 
affecting the polynomial coefficients in
(\ref{eq:tumapp3}) and (\ref{eq:tumapp4}).

The Cartesian coordinate system $(x,y,z)$ rotates with
respect to the star as $\bfw$ changes orientation
under the action of ${\bf N}$.
Among other things, this causes $\alpha$ to change
with time.
In order to write down Euler's equations,
it is necessary to reexpress ${\bf N}$ in body coordinates
that are fixed with respect to the star.
We choose the body axes to be the principal axes of
the star's ellipsoid of inertia,
${\bf e}_1$, ${\bf e}_2$ and ${\bf e}_3$;
the magnetic axis ${\bf m}$, which is also fixed with
respect to the star, is taken to lie in the plane
spanned by ${\bf e}_1$ and ${\bf e}_3$, at an angle
$\chi$ to ${\bf e}_3$ (a slight loss of generality).
The transformation from $(x,y,z)$ to body coordinates
is time-dependent, but it is not an Euler
transformation because $(x,y,z)$ is a non-inertial
frame. It is described by a matrix $[A_{ij}]$,
defined through 
$({\bf e}_1,{\bf e}_2,{\bf e}_3)=
[A_{ij}]\cdot ({\bf i},{\bf j}, {\bf k})$,
which can be represented in the form
\begin{equation}
 [A_{ij}] =
 \left(
  \begin{array}{ccc}
  \cos\phi\cos\theta\cos\psi - \sin\phi\sin\psi & 
  \sin\phi\cos\theta\cos\psi + \cos\phi\sin\psi & 
  - \sin\theta\cos\psi \\ 
  - \cos\phi\cos\theta\sin\psi - \sin\phi\cos\psi & 
  - \sin\phi\cos\theta\sin\psi + \cos\phi\cos\psi & 
  \sin\theta\sin\psi \\ 
  \cos\phi\sin\theta & \sin\phi\sin\theta & \cos\theta
  \end{array}
 \right),
 \label{eq:tumapp5}
\end{equation}
where the angles $\phi$, $\psi$, $\theta$
({\em not} Euler angles) depend on $t$ through the
principal angular velocity components $\omega_1$,
$\omega_2$, $\omega_3$ as follows:
\begin{eqnarray}
 \cos\phi 
 & = &
 \frac{\omega\cos\chi - \omega_3\cos\alpha}
  {\sin\alpha(\omega^2-\omega_3^2)^{1/2}},
 \label{eq:tumapp6} \\
 \sin\phi 
 & = &
 - \frac{\omega_2\sin\chi}
  {\sin\alpha(\omega^2-\omega_3^2)^{1/2}},
 \label{eq:tumapp7} \\
 \cos\psi 
 & = &
 - \frac{\omega_1}
  {(\omega^2-\omega_3^2)^{1/2}},
 \label{eq:tumapp8} \\
 \sin\psi 
 & = &
 \frac{\omega_2}
  {(\omega^2-\omega_3^2)^{1/2}},
 \label{eq:tumapp9} \\
 \cos\theta 
 & = &
 \omega_3/\omega,
 \label{eq:tumapp10} \\
 \sin\theta 
 & = &
 (1-\omega_3^2/\omega^2)^{1/2}.
 \label{eq:tumapp11}
\end{eqnarray}
The angle $\alpha$ is chosen to lie in the range
$0\leq \alpha\leq \pi$ and satisfies
\begin{equation}
 \omega\cos\alpha
 =
 \omega_1 \sin\chi + \omega_3 \cos\chi.
 \label{eq:tumapp12}
\end{equation}
Upon substituting (\ref{eq:tumapp6}) to (\ref{eq:tumapp12})
into (\ref{eq:tumapp5}) and employing the relation
$(N_1,N_2,N_3)=[A_{ij}]\cdot(N_x,N_y,N_z)$,
we arrive at the principal components of the
radiation torque featured
on the right-hand sides of Euler's equations
(\ref{eq:tum2}), (\ref{eq:tum3}) and (\ref{eq:tum4}).

\section{APPROXIMATE ANALYTIC SOLUTION FOR $\epsilon'=0$,
 $\chi=90\arcdeg$ 
 \label{sec:tumappb}}
In this appendix, we derive an approximate solution to
Euler's equations (\ref{eq:tum2}), (\ref{eq:tum3})
and (\ref{eq:tum4})
for a biaxial star ($\epsilon'=0$) with $\chi=90\arcdeg$. 
The solution is
accurate provided the precession periods $\tau_1$
(radiative) and $\tau_2$ (free) are much smaller than
the braking time-scale $\tau_0$.

When the slow braking terms, proportional to
$(\omega_0\tau_0)^{-1}F(x_0)$,
are neglected relative to
the fast precessive terms, proportional to
$\epsilon$ and $(\omega_0\tau_0)^{-1}uG(x_0)$,
Euler's equations reduce to 
the zeroth-order system (\ref{eq:tum18})
to (\ref{eq:tum20}) which, in the special case
$\chi=90\arcdeg$, reduces to
\begin{eqnarray}
 \dot{u}_1
 & = &
 -\epsilon u_2 u_3~,
 \label{eq:tumapp13}
 \\
 \dot{u}_2
 & = &
 \epsilon (1 + \beta^{-1}) u_1 u_3~,
 \label{eq:tumapp14}
 \\
 \dot{u}_3
 & = &
 -\epsilon \beta^{-1} u_1 u_2~.
 \label{eq:tumapp15}
\end{eqnarray}
The solutions of (\ref{eq:tumapp13}) to (\ref{eq:tumapp15})
are Jacobian elliptic functions. The relative signs of
the coefficients on the right-hand sides
determine the solution branch, given the physical
requirement that $u_1$, $u_2$ and $u_3$ are
real quantities whose squared amplitudes are non-negative.
We distinguish three solution branches:

\noindent Case I: $\beta>0$.
\begin{eqnarray}
 u_1 
 & = &
 \left(
  \frac{\gamma+\eta\beta}{\beta+1}
 \right)^{1/2}
 {\rm cn}
 \left[ \epsilon
  \left( \frac{\eta-\gamma}{\beta} \right)^{1/2} s
  + \Phi
 \right],
 \label{eq:tumapp16}
 \\
 u_2 
 & = &
 \left(
  \frac{\gamma+\eta\beta}{\beta}
 \right)^{1/2}
 {\rm sn}
 \left[ \epsilon
  \left( \frac{\eta-\gamma}{\beta} \right)^{1/2} s
  + \Phi
 \right],
 \label{eq:tumapp17}
 \\
 u_3 
 & = &
 \left(
  \frac{\eta-\gamma}{\beta+1}
 \right)^{1/2}
 {\rm dn}
 \left[ \epsilon
  \left( \frac{\eta-\gamma}{\beta} \right)^{1/2} s
  + \Phi
 \right],
 \label{eq:tumapp18}
 \\
 k^2
 & = &
 \frac{\gamma+\eta\beta}{\beta(\eta-\gamma)}~.
 \label{eq:tumapp19}
\end{eqnarray}

\noindent Case II: $-1<\beta<0$.
\begin{eqnarray}
 u_1 
 & = &
 \gamma^{1/2}
 {\rm cn}
 \left[ \epsilon
  \left( \frac{\eta-\gamma}{-\beta} \right)^{1/2} s
  + \Phi
 \right],
 \label{eq:tumapp20}
 \\
 u_2 
 & = &
 (\eta-\gamma)^{1/2}
 {\rm dn}
 \left[ \epsilon
  \left( \frac{\eta-\gamma}{-\beta} \right)^{1/2} s
  + \Phi
 \right],
 \label{eq:tumapp21}
 \\
 u_3 
 & = &
 \left(
  \frac{\gamma}{-\beta}
 \right)^{1/2}
 {\rm sn}
 \left[ \epsilon
  \left( \frac{\eta-\gamma}{-\beta} \right)^{1/2} s
  + \Phi
 \right],
 \label{eq:tumapp22}
 \\
 k^2
 & = &
 - \frac{(\beta+1)\gamma}{\beta(\eta-\gamma)}~.
 \label{eq:tumapp23}
\end{eqnarray}

\noindent Case III: $\beta<-1$.
\begin{eqnarray}
 u_1 
 & = &
 \left(
  \frac{\gamma+\eta\beta}{\beta+1}
 \right)^{1/2}
 {\rm sn}
 \left[ -\epsilon
  \left( \frac{\beta+1}{\beta} \right)^{1/2} 
  \left( \frac{\gamma}{-\beta} \right)^{1/2} s
  + \Phi
 \right],
 \label{eq:tumapp24}
 \\
 u_2 
 & = &
 \left(
  \frac{\gamma+\eta\beta}{\beta}
 \right)^{1/2}
 {\rm cn}
 \left[ -\epsilon
  \left( \frac{\beta+1}{\beta} \right)^{1/2}
  \left( \frac{\gamma}{-\beta} \right)^{1/2} s
  + \Phi
 \right],
 \label{eq:tumapp25}
 \\
 u_3 
 & = &
 \left(
  \frac{\gamma}{-\beta}
 \right)^{1/2}
 {\rm dn}
 \left[ -\epsilon
  \left( \frac{\beta+1}{\beta} \right)^{1/2}
  \left( \frac{\gamma}{-\beta} \right)^{1/2} s
  + \Phi
 \right],
 \label{eq:tumapp26}
 \\
 k^2
 & = &
 \frac{\gamma+\eta\beta}{(\beta+1)\gamma}~.
 \label{eq:tumapp27}
\end{eqnarray}
In (\ref{eq:tumapp16}) to (\ref{eq:tumapp27}),
the quantities $\eta$, $\gamma$ and $\Phi$ are all
constants of the motion (see Section \ref{sec:tum2c}).
The trivial cases $\beta=-1$ 
(harmonic precession about ${\bf e}_2$)
and $\beta=0$ (spherical star; see
\markcite{dav70}Davis \& Goldstein 1970) are not treated
here.

When terms proportional to $(\omega_0\tau_0)^{-1}F(x_0)$
are restored to Euler's equations, the above solutions
remain approximately valid, but the constants of the
motion are converted into slowly varying functions of $s$.
We now compute the slow evolution of $\eta(s)$,
$\gamma(s)$ and $\Phi(s)$. First, we substitute
(\ref{eq:tumapp16}) to (\ref{eq:tumapp27})
into 
(\ref{eq:tum2}), (\ref{eq:tum3}) and (\ref{eq:tum4}) 
and perform 
the time derivatives explicitly. For each solution
branch, this results in a system of three equations
linear in $\dot{\eta}$, $\dot{\gamma}$ and $\dot{\Phi}$.
For example, the $\beta>0$ branch yields
\begin{equation}
 \frac{\dot{\gamma}+\dot{\eta}\beta}
  {2(\gamma+\eta\beta)}
 =
 -\frac{F(x_0)\eta}{\omega_0\tau_0}
 \, {\rm sn}^2~,
 \label{eq:tumapp28}
\end{equation}
\begin{equation}
 \left[ \epsilon 
  \left( \frac{\beta}{\eta-\gamma} \right)^{1/2}
  \frac{\dot{\eta}-\dot{\gamma}}{2\beta} s
  + \dot{\Phi}
 \right] {\rm dn}^2
 =
 -\frac{F(x_0)\eta}{\omega_0\tau_0}
 \,{\rm sn}\,{\rm cn}\,{\rm dn}~,
 \label{eq:tumapp29}
\end{equation}
\begin{equation}
 - \frac{\dot{\gamma}+\dot{\eta}\beta}
  {2\beta (\eta-\gamma)}
 \,{\rm cn}^2
 +
 \frac{\dot{\eta}-\dot{\gamma}}
  {2 (\eta-\gamma)}
 \,{\rm dn}^2
 =
 - \frac{F(x_0)\eta}{(1+\epsilon)\omega_0\tau_0}
 \,{\rm dn}^2
 +
 \frac{\epsilon^2(\gamma+\eta\beta)}
  {(1+\epsilon)\beta^{3/2}(\eta-\gamma)^{1/2}}
 \,{\rm sn}\,{\rm cn}\,{\rm dn}~.
 \label{eq:tumapp30}
\end{equation}
In (\ref{eq:tumapp28}) to (\ref{eq:tumapp30}),
Euler's equations are linearly combined in such
a way that the coefficients of
$\dot{\eta}$, $\dot{\gamma}$ and $\dot{\Phi}$
are squares of elliptic functions, in order
to ensure a non-trivial result when averaging
over $\Phi$.
The (omitted) arguments of the elliptic functions 
are the same as in
(\ref{eq:tumapp16}) to (\ref{eq:tumapp18}).
We now average over $\Phi$, holding constant
$\eta$, $\gamma$, 
and their derivatives.
The results for the three solution branches
are as follows:

\noindent Case I: $\beta>0$.
\begin{eqnarray}
 \dot{\eta}
 & = &
 -\frac{2 F(x_0) \eta [ (1+\epsilon)(\gamma+\eta\beta) I_1
   (I_2+\beta I_3) + \beta(\eta-\gamma) I_3 ]}
 {\omega_0\tau_0(1+\epsilon)\beta(\beta+1) I_3}~,
 \label{eq:tumapp31}
 \\
 \dot{\gamma}
 & = &
 \frac{2 F(x_0) \eta [ (1+\epsilon)(\gamma+\eta\beta) I_1
   (I_2 - I_3) + \beta(\eta-\gamma) I_3 ]}
 {\omega_0\tau_0(1+\epsilon)(\beta+1) I_3}~,
 \label{eq:tumapp32}
 \\
 \dot{\Phi}
 & = &
 \frac{F(x_0) \eta [ (1+\epsilon)(\gamma+\eta\beta) I_1
   I_2 + \beta(\eta-\gamma) I_3 ]}
 {\omega_0\tau_0(1+\epsilon)\beta^2 I_3}
 \left(
  \frac{\beta}{\eta-\gamma} 
 \right)^{1/2} \epsilon s~.
 \label{eq:tumapp33}
\end{eqnarray}

\noindent Case II: $-1<\beta<0$.
\begin{eqnarray}
 \dot{\eta}
 & = &
 -\frac{2 F(x_0) \eta \{ (1+\epsilon)\beta(\eta-\gamma) I_3
   + \gamma I_1 [ \beta I_3 - (\beta+1) I_2 ] \} }
 {\omega_0\tau_0(1+\epsilon)\beta I_3}~,
 \label{eq:tumapp34}
 \\
 \dot{\gamma}
 & = &
 -\frac{2 F(x_0) \eta\gamma I_1}
 {\omega_0\tau_0(1+\epsilon)}~,
 \label{eq:tumapp35}
 \\
 \dot{\Phi}
 & = &
 \frac{\epsilon^2 I_4}{(1+\epsilon) I_3}
 \left(
  \frac{\eta-\gamma}{-\beta}
 \right)^{1/2}
 -\frac{F(x_0) \eta [ (1+\epsilon)\beta(\eta-\gamma) I_3
   - (\beta+1)\gamma I_1 I_2  ] }
 {\omega_0\tau_0(1+\epsilon)\beta^2 I_3}
 \left(
  \frac{-\beta}{\eta-\gamma} 
 \right)^{1/2} \epsilon s~.
 \phantom{XX}
 \label{eq:tumapp36}
\end{eqnarray}

\noindent Case III: $\beta<-1$.
\begin{eqnarray}
 \dot{\eta}
 & = &
 -\frac{2 F(x_0) \eta \{ (1+\epsilon)(\gamma+\eta\beta) I_2
   [ (\beta+1) I_3 + I_1 ] - (\beta+1)\gamma I_3 \} }
 {\omega_0\tau_0(1+\epsilon)\beta(\beta+1) I_3}~,
 \label{eq:tumapp37}
 \\
 \dot{\gamma}
 & = &
 -\frac{2 F(x_0) \eta [ -(1+\epsilon)(\gamma+\eta\beta)
   I_1 I_2 + (\beta+1)\gamma I_3 ] }
 {\omega_0\tau_0(1+\epsilon)(\beta+1) I_3}~,
 \label{eq:tumapp38}
 \\
 \dot{\Phi}
 & = &
 \frac{F(x_0) \eta [ -(1+\epsilon)(\gamma+\eta\beta)
   I_1 I_2 + (\beta+1)\gamma I_3 ] }
 {\omega_0\tau_0(1+\epsilon)\beta (\beta+1) I_3}
 \left(
  \frac{\beta+1}{-\gamma} 
 \right)^{1/2} \epsilon s~.
 \label{eq:tumapp39}
\end{eqnarray}
In (\ref{eq:tumapp31}) to (\ref{eq:tumapp39}), the $\Phi$ averages
$I_i$ ($1\leq i\leq 4$) are all functions of the modulus $k$
appropriate for each branch and are defined by
$I_1=\langle {\rm sn}^2\Phi \rangle$,
$I_2=\langle {\rm cn}^2\Phi \rangle$,
$I_3=\langle {\rm dn}^2\Phi \rangle$,
and
$I_4=\langle {\rm cn}^2\Phi\,{\rm dn}^2\Phi \rangle$.

In the limit $\beta\rightarrow\infty$
(i.e.\ $k\rightarrow 0$), 
the three solution branches merge together into one,
the elliptic functions reduce to trigonometric functions,
and the 
equations for $\dot{\eta}$, $\dot{\gamma}$, $\dot{\Phi}$
reduce to those given by
\markcite{gol70}Goldreich (1970).



\clearpage

\end{document}